\documentclass[reprint,aps,pre,longbibliography,amsmath,amssymb]{revtex4-1}
\usepackage{graphicx}
\usepackage{hyperref}

\newcommand{\va}{c}    
\newcommand{\wn}{k}    
\newcommand{\Aff}{{\rm Aff}(1,\mathbb{R})}
\newcommand{\s}[1]{#1_0}
\newcommand{\wb}{D}    
\newcommand{\xb}{d}
\renewcommand{\wp}{P}  
\newcommand{\xp}{p}
\newcommand{\wg}{A}    
\newcommand{\sg}{\s{A}}
\newcommand{\xg}{a}
\newcommand{\ws}{S}    
\renewcommand{\ss}{\s{S}}
\newcommand{\xs}{s}
\newcommand{\wa}{R}    
\newcommand{\xa}{r}
\newcommand{\wt}{B}    
\newcommand{\st}{\s{B}}
\newcommand{\xt}{b}
\newcommand{\V}{\Omega}

\begin{document}

\title{A scale-invariant model of marine population dynamics}
\author{Jos\'{e} A. Capit\'{a}n}
\email{jcapitan@math.uc3m.es}
\affiliation{Grupo Interdisciplinar de Sistemas Complejos (GISC), Departamento de Matem\'aticas,\\
Escuela Politécnica Superior, Universidad Carlos III de Madrid, E28911 Leganés, Spain}
\author{Gustav W. Delius}
\email{gwd2@york.ac.uk}
\affiliation{Department of Mathematics, University of York, York, UK}

\begin{abstract}

A striking feature of the marine ecosystem is the regularity in its
size spectrum: the abundance of organisms as a function of their
weight approximately follows a power law over almost ten orders of magnitude. We
interpret this as evidence that the population dynamics in the ocean
is approximately scale-invariant. We use this invariance in the construction and
solution of a size-structured dynamical population model.

Starting from a Markov model encoding the basic
processes of predation, reproduction, maintenance respiration and
intrinsic mortality, we derive a partial integro-differential
equation describing the dependence of abundance on weight and time.
Our model represents an extension of the jump-growth model and hence
also of earlier models based on the McKendrick--von Foerster equation.
The model is scale-invariant provided the rate functions of the
stochastic processes have certain scaling properties.

We determine the steady-state power law solution, whose exponent is determined by the relative scaling
between the rates of the density-dependent processes (predation) and the rates of the
density-independent processes (reproduction, maintenance, mortality).
We study the stability of the steady-state against small perturbations and find that
inclusion of maintenance respiration and reproduction in the model has a
strong stabilising effect. Furthermore, the steady state
is unstable against a change in the overall population density unless
the reproduction rate exceeds a certain threshold.

\end{abstract}


\maketitle

\section{\label{s:intro}Introduction}

The population dynamics in the pelagic marine ecosystem (the open sea) are particularly amenable to mathematical modelling and analytic understanding. Because size is the most
important factor in determining who eats who, rather than species \cite{Jennings2001}, it is possible to work with a model in which a single function $\phi(w,t)$ describes the
total population density of organisms of weight $w$ at time $t$, aggregated over all species. We can build on many previous works using partial integro-differential equations to
describe the time evolution of this total population density, in particular \cite{SilvertPlatt1978,SilvertPlatt1980,Cushing1992,CamachoSole2001,BenoitRochet2004,Arino2004,Datta2008,Blanchard2009}.

In this paper we exploit an additional special property of the pelagic zone: its approximate physical scale invariance. The open sea looks similar at a range of scales. That is due to the fact that
over many orders of magnitude there are no physical features and no strong physical principles that would single out a particular intermediate scale. We expect that this approximate
scale invariance of the environment breaks down only at small scales at which molecular diffusion processes dominate  and at large scale at which geography affects ocean
currents. This kind of scale-invariance is not present in terrestrial environments,
where not only the physical structure sets a scale, but where in addition the effects of gravity quickly become important for larger
organisms.

It is not a priori guaranteed that the scale invariance of the physical environment will also lead to scale invariance of the ecosystem. The organisms populating the environment
and their interactions could break the scale invariance. This paper is however based on the assumption that evolution, in its drive to make use of all available ecological niches,
has made optimal use of the scale invariant environment by filling it with an ecosystem that roughly preserves scale invariance.
It is beyond the scope of the current paper to investigate the evolutionary mechanisms that would lead to the self-organisation of such a scale invariant ecosystem. We will be
content with deriving the consequences of scale invariance for the dynamics of the population density.

The main observational evidence for approximate scale invariance is that the equilibrium size distribution of organisms in the open ocean is approximately given by a power law
\(\phi(w)\propto w^{-\gamma}\), valid over almost ten orders of magnitude \cite{Sheldon1972,Boudreau1992,KerrDickie2001,JenningsMackinson2003,Quinones2003,Marquet2005}. Several theoretical models have been proposed in
the past to derive this steady-state size spectrum \cite{PlattDenman1977,SilvertPlatt1978,SilvertPlatt1980,CamachoSole2001,BenoitRochet2004,Arino2004,Law2009,Datta2008} but, as far as we know, in this paper we
are the first to point out that this can be understood as a consequence of the scale invariance of the underlying dynamical model.

The main processes that affect the abundance of organisms as a function of weight are predation, reproduction, maintenance respiration and intrinsic death. We will in the next section write down a dynamical model that incorporates all these processes. In general such a model will not reproduce the observed power-law size-spectrum in the steady state. This is the reason why existing work either only models predation \cite{SilvertPlatt1978,SilvertPlatt1980,BenoitRochet2004,Arino2004,Datta2008} or makes the simplifying assumption that reproduction, maintenance respiration and death are exactly proportional to predation \cite{CamachoSole2001}. We are able to avoid those difficulties by exploiting scale invariance.

While scale invariance leads to a power-law steady-state solution, it does not guarantee that this steady state is a stable equilibrium solution. Indeed, a linear stability analysis
of the jump-growth model \cite{Datta2010}, which is also scale invariant, showed that, for realistic choices of the parameters, the steady state in that model is unstable against
small perturbations. The steady state observed in nature is stable, so the pure jump-growth model is missing some important stabilising effect. This was the motivation for the
investigations of the more general model in this paper. We will see that the inclusion of maintenance respiration and reproduction has a strong stabilising effect.

We do not model spatial variation in the population density. Nevertheless our model formally resembles a model in one space and one time dimension,
where the space dimension represents the weight of the organisms. This formal analogy between space and weight will become particularly clear when we transform from the weight
variable $w$ to the logarithmic weight variable $x$ in Section~\ref{s:log}. A scale transformation in $w$ then corresponds to a translation in $x$. In the special case where the
scale transformations do not affect the time variable we end up with a model that is translationally invariant in both space and time, and many of the usual techniques
can be applied, like for example the use of the standard Fourier transform in the solution of the linearised model in Section~\ref{s:stability}.

Another difference between our model and those usually studied by physicists derives from the fact that feeding is a non-local interaction in weight space. Most fish and plankton do not feed on other individuals that are close to their own weight. Instead they prefer prey that are substantially smaller. Similarly they produce offspring at a weight far below their own. This is the reason
why the population density $\phi(w,t)$ is modelled not by a partial differential equation but by the partial integro-differential equation \eqref{eq:detcon} involving integral terms
that encode the feeding behaviour \eqref{eq:detconpred} and reproductive behaviour \eqref{eq:detconrep}. In the linear stability analysis this leads to the non-local dispersion
relation \eqref{eq:lamp}.

The organisation of this paper is as follows. In Section~\ref{s:det} we formulate a stochastic model encoding feeding, reproduction, maintenance, and death. We do not make any
assumptions about the rates for these processes but keep them as general parameters of the model. By taking the macroscopic limit of the stochastic model we derive the deterministic
evolution equation \eqref{eq:detcon}. The derivation of a Fokker--Planck equation for the stochastic fluctuations away from
the macroscopic model is left to Appendix~\ref{s:master}. In Section~\ref{s:scale} we impose scale-invariance and derive sufficient scaling conditions \eqref{eq:rtx} for the rate functions. Section
\ref{s:log} is devoted
to transforming the evolution equation into the more convenient form \eqref{eq:logtrans}. We then use scale invariance to find a power-law steady state solution of the model
in Section~\ref{s:ss},
where we discuss as well its main properties and restrictions. The overall population level is predicted by the model and we show in Appendix~\ref{s:uzero} that is positive under
biologically relevant assumptions. A time-damped power-law solution can also be found, as we show in Appendix~\ref{s:sst}. In order to investigate the
stability of the steady state we determine the
spectrum of small perturbations around the steady state in Section~\ref{s:stability}. We summarise our results in the Section~\ref{s:discussion}. In Appendix~\ref{s:absorption} we describe how variability in the absorption efficiency can be approximated by
the addition of a diffusion term to the model.

\section{\label{s:det}Size-structured population model}

We start by constructing a stochastic model for the dependence of abundance of individuals on weight and time, taking into account the basic processes of predation,
reproduction, maintenance and intrinsic mortality.

As in previous work \cite{Datta2008}, instead of keeping track of the weight of each individual, we aggregate individuals of similar weight into discretised weight brackets
$[w_i,w_{i+1})$ for $i \in \mathbb{Z}$. Weight is the only attribute of individuals that is used in this model. Species identity and life stage are ignored.
The weight distribution of individuals in a large fixed volume $\V$ is described
by the vector ${\bf n} = (\dots,n_{-1},n_0,n_1,\dots)$ whose entries give the number of organisms in each weight bracket. We will later let the size of these brackets go to zero to
obtain the continuum model. Working with discrete weight brackets allows us to avoid the mathematical complexities involved when trying to describe a stochastic model
directly in continuous space, see for example \cite{Kondratiev2008}.

\begin{figure}
\includegraphics[width=8.5cm]{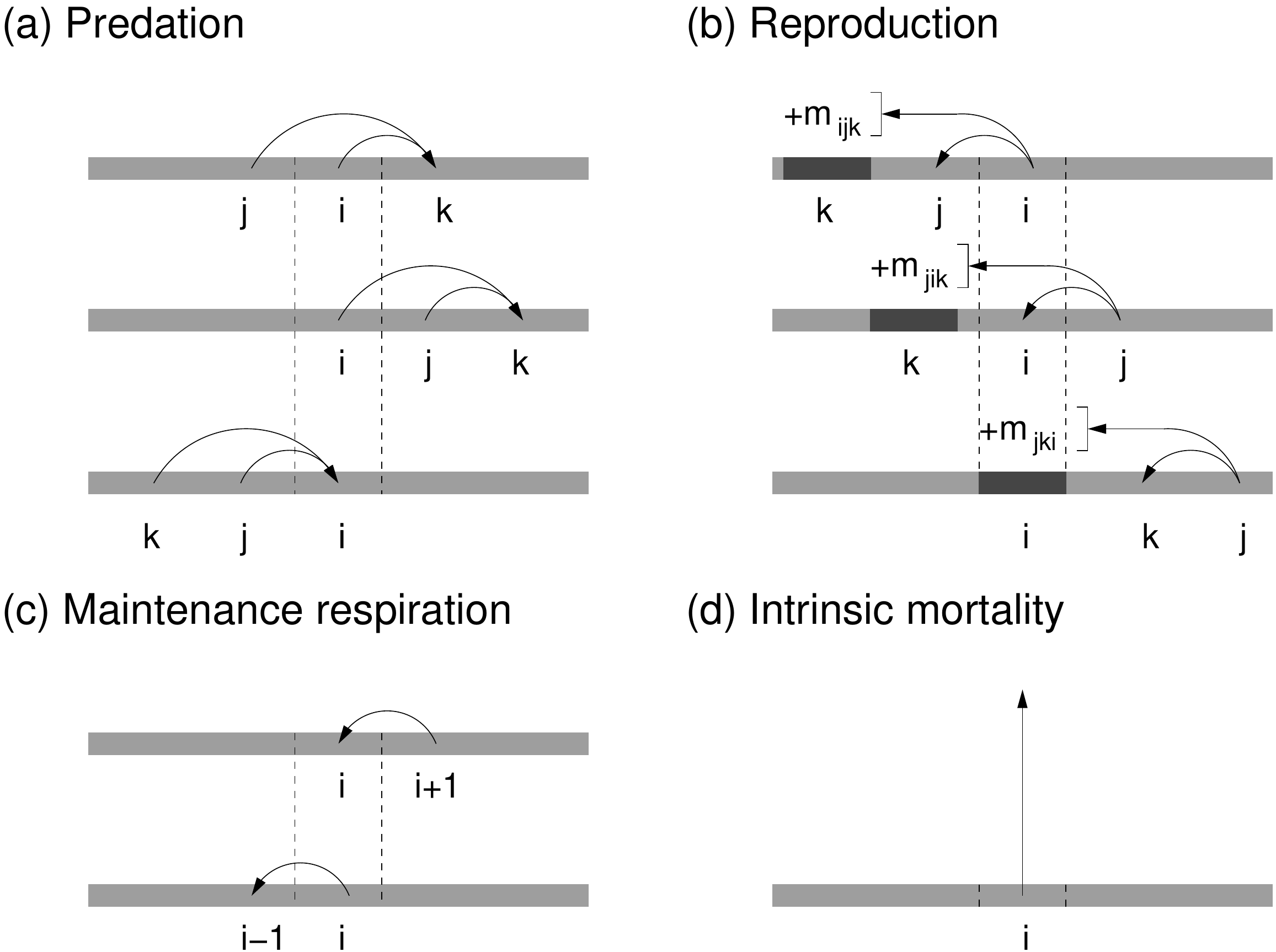}
\caption{\label{fig:proc}Individual stochastic processes involved in the model. The figure shows all the events that affect the number of organisms in weight bracket $i$.
The arrows indicate the movement of individuals between weight brackets. (a) Predation: There are three predation processes affecting bracket $i$. The first represents a predator
in $i$ eating a prey in $j$ and absorbing enough prey weight to end up in $k$. The second represents a prey in $i$ being eaten and the third represents a predator entering $i$ after
feeding. (b) Reproduction: in the first process an individual in $i$ produces $m_{ijk}$ offspring in bracket $k$ and, as a consequence, decreases its weight to $j$. In addition, there
are two processes that increase the number of individuals in $i$. (c) Maintenance respiration: these processes move individuals to the next-lower bracket. There are two ways in which
this affects to bracket $i$. (d) Intrinsic mortality: with a certain rate, a single individual in bracket $i$ is removed from the system.}
\end{figure}

The primary stochastic processes involved in the model are illustrated in Figure~\ref{fig:proc}, which shows the various ways in which an event can affect the number of organisms
in bracket $i$. Below we will show how the deterministic equation can be read off directly from this figure. In Appendix~\ref{s:master} we start from a master equation
for the probability distribution $P({\bf n},t)$ and carry out the systematic expansion of van Kampen \cite{vanKampen} in powers of the inverse system volume $\V^{-1}$.
The same method was used in \cite{Datta2008} to derive a jump-growth equation.
This method is based on splitting each variable $n_i(t)$ into a deterministic, macroscopic component $\phi_i(t)$ describing the density of individuals in weight
bracket $i$, and a fluctuation component $\eta_i(t)$ as
\begin{equation}\label{eq:kampen}
n_i(t) = \V\phi_i(t)+\V^{1/2}\eta_i(t).
\end{equation}
The powers of volume are chosen so that the new variables $\phi_i$ and $\eta_i$ no longer scale with the system volume.
This method not only gives the macroscopic behaviour for the densities $\phi_i(t)$ at leading order in the expansion, at higher orders it describes the stochastic fluctuations around the macroscopic solution
as well, giving at next-to-leading order a linear Fokker--Planck equation.
However, because the system volume is so large, in this paper we will concentrate on the macroscopic, deterministic equation.

From Figure~\ref{fig:proc} we can obtain the contributions to the time evolution of $\phi_i$ from each of the processes we consider: predation (${\rm\wp}$), reproduction (${\rm\wt}$ for Birth), maintenance respiration (${\rm\wa}$ for Respiration) and intrinsic mortality (${\rm\wb}$ for Death),
\begin{equation}\label{eq:det}
\frac{d\phi_i}{d t}=\left(\frac{d\phi_i}{d t}\right)_{\rm\wp}
+\left(\frac{d\phi_i}{d t}\right)_{\rm\wt}+\left(\frac{d\phi_i}{d t}\right)_{\rm\wa}
+\left(\frac{d\phi_i}{d t}\right)_{\rm\wb}.
\end{equation}
We will describe each of them in turn below.

\subsection{Predation}

A predation event moves a predator from a weight bracket $i$ before feeding to a higher weight bracket $k$ after feeding and removes a prey organism
from a weight bracket $j$. Let $\wp_{ijk}$ be the rate constant for such a predation event. As illustrated in Figure~\ref{fig:proc}a, there are three ways in which a predation event
can affect the number $n_i$ of individuals in a weight bracket $i$: 1) an organism in bracket $i$ can eat another organism and grow; 2) an organism in bracket $i$ can be eaten by
another organism; and 3) an individual in a lower bracket can absorb enough prey weight and grow into bracket $i$. Because the rate at which a particular individual will encounter
prey will be proportional
to the density of prey, the probability that one of the $n_i$ individuals in bracket $i$ eats any of the $n_j$ individuals of bracket $j$ and increases its size to bracket $k$ in
the time interval from $t$ to $t+dt$ is $\wp_{ijk} n_i n_j\V^{-1}dt$. Hence the contribution of the predation events to the deterministic time evolution of the density $\phi_i$ of organisms in bracket $i$ is
\begin{equation}\label{eq:deterpred}
\left(\frac{d\phi_i}{dt}\right)_{\rm\wp}=\sum_{j,k}(-\wp_{ijk}\phi_i\phi_j-\wp_{jik}\phi_j\phi_i+\wp_{jki}\phi_j\phi_k).
\end{equation}

Our model for predation can be viewed as a generalisation of the Smoluchowski coagulation equation \cite{Smoluchowski1916,Drake1972} which is obtained in the special case when the resulting weight $w_k$ is equal to the sum of the weights $w_i$ and $w_j$ of predator and prey.

\subsection{Reproduction}

Most fish reproduce by laying a large number of eggs that are subject to heavy predation. Only a small fraction of the eggs survives to hatch and join the consumer spectrum.
In a size-spectrum model however, in which size is the only attribute of an organism and its life stage is  ignored,  all  organisms  are  assumed  to  be  prey  and predator simultaneously from the moment they are born.  We can therefore not model the egg life stage and therefore can not provide an entirely realistic model of the reproductive processes. Our model is thus designed to only capture two features of reproduction that we deem essential to the size-spectrum dynamics, namely that it moves biomass up the size spectrum from large weight to small weight and that it replenishes the population numbers at smaller weight.

We assume that a reproduction event moves a parent organism from a weight
bracket $i$ to a lower weight bracket $j$ and produces a large number $m_{ijk}$ of smaller offspring in weight bracket $k$. We set the number of offspring to
$m_{ijk}=\lfloor(w_i-w_j)/w_k\rfloor$, where $\lfloor x\rfloor$ denotes the largest integer smaller than $x$, so that their combined weight is approximately equal to the weight
lost by the parent.
Let $\wt_{ijk}$ be the rate constant for such an event. The probability that one of the $n_i$ individuals in bracket $i$ reproduces in such an event in $dt$ is then $\wt_{ijk}n_i dt$
(note that reproduction, unlike predation, is a density-independent process).

As depicted in Figure~\ref{fig:proc}b, there are three
ways in which reproduction changes the number of the organisms in bracket $i$: 1) the parent belongs to the $i$-th bracket, and after spawning moves to a lower bracket;
2) a parent looses weight during spawning and moves to bracket $i$; and 3) the bracket $i$ receives the offspring from a reproduction event.
The contribution of these three possibilities to the deterministic equation is therefore given by
\begin{equation}\label{eq:deterrep}
\left(\frac{d\phi_i}{dt}\right)_{\rm\wt}=\sum_{j,k}\left(-\wt_{ijk}\phi_i+\wt_{jik}\phi_j+m_{jki}\wt_{jki}\phi_j\right).
\end{equation}

\subsection{Maintenance and mortality}

In between feeding events, organisms continuously draw upon their reserves to maintain themselves. The weight loss due to maintenance respiration is modelled by events that move individuals to the next-lower weight bracket, assuming that the width of each interval is small enough (this is not a
restriction, because at the end we will take the continuum limit where all these widths tend to zero). The probability that any of the $n_i$ individuals in
the bracket $i$ undergoes such an event in $dt$ is $\wa_i n_i dt$, where $\wa_i$ is the maintenance respiration rate for the bracket $i$. Figure~\ref{fig:proc}c shows the two ways
that these primary processes change the number of individuals in bracket $i$, giving the contribution
\begin{equation}\label{eq:detmain}
\left(\frac{d\phi_i}{dt}\right)_{\rm\wa}=\wa_{i+1}\phi_{i+1}-\wa_i\phi_i.
\end{equation}

Organisms can also die for reasons other than being eaten. This introduces a fourth process in the model that accounts for intrinsic mortality. With a probability
$\wb_i n_i dt$, a single individual in bracket $i$ is removed from the system in $dt$ (see Figure~\ref{fig:proc}d). This gives the contribution
\begin{equation}\label{eq:detmor}
\left(\frac{d\phi_i}{dt}\right)_{\rm\wb}=-\wb_i\phi_i.
\end{equation}
Note that, like reproduction, both maintenance and intrinsic mortality are modelled as density-independent processes.

\subsection{Continuum limit}

We now take the continuum limit of each contribution to the macroscopic equation by writing $\Delta_i=w_{i+1}-w_i$ and then taking the limit $\Delta_i\rightarrow 0$ uniformly for
all $i$. The variables $\phi_i(t)$ are combined into a density $\phi(w,t)$ of individuals per unit weight per unit volume as a function of weight and time so that
$\phi(w_i,t)=\phi_i(t)/\Delta_i$. The sum over weight brackets is replaced by an integral, $\sum_i\Delta_i\rightarrow \int dw$.
Continuum rate functions are introduced as
\begin{equation}
\begin{split}
\wp(w_i,w_j,w_k)&=\wp_{ijk}/\Delta_k,\\
\wt(w_i,w_j,w_k)&=\wt_{ijk}/(\Delta_j\Delta_k),\\
\wa(w_i)&=\Delta_i \wa_i,\\
\wb(w_i)&=\wb_i.
\end{split}
\end{equation}
After taking the continuum limit, the ordinary differential equations \eqref{eq:det} for the $\phi_i$ assemble into a partial differential equation for $\phi(w,t)$,
\begin{widetext}
\begin{equation}\label{eq:detcon}
\frac{\partial\phi(w,t)}{\partial t}=\left(\frac{\partial\phi(w,t)}{\partial t}\right)_{\rm\wp}
+\left(\frac{\partial\phi(w,t)}{\partial t}\right)_{\rm\wt}+\left(\frac{\partial\phi(w,t)}{\partial t}\right)_{\rm\wa}
+\left(\frac{\partial\phi(w,t)}{\partial t}\right)_{\rm\wb},
\end{equation}
which will be the object of study in the remainder of the paper.
The contribution from predation \eqref{eq:deterpred} in this limit becomes
\begin{equation}\label{eq:detconpred}
\left(\frac{\partial\phi(w,t)}{\partial t}\right)_{\rm\wp}=
\int dw'\int dw'' \left(-[\wp(w,w',w'')+\wp(w',w,w'')]\phi(w,t)\phi(w',t)
+\wp(w',w'',w)\phi(w',t)\phi(w'',t)\right).
\end{equation}
Integrals run over all positive weights. Similarly, the continuum limit of \eqref{eq:deterrep} is
\begin{equation}\label{eq:detconrep}
\left(\frac{\partial\phi(w,t)}{\partial t}\right)_{\rm\wt}=
\int dw'\int dw''\left(-\wt(w,w',w'')\phi(w,t)+\wt(w',w,w'')\phi(w',t)
+\frac{w'-w''}{w}\wt(w',w'',w)\phi(w',t)\right).
\end{equation}
\end{widetext}
Maintenance and intrinsic mortality are described by
\begin{equation}\label{eq:detconmain}
\left(\frac{\partial\phi(w,t)}{\partial t}\right)_{\rm\wa}=
\frac{\partial}{\partial w}\left[\wa(w)\phi(w,t)\right]
\end{equation}
and
\begin{equation}\label{eq:detconmor}
\left(\frac{\partial\phi(w,t)}{\partial t}\right)_{\rm\wb}=-\wb(w)\phi(w,t),
\end{equation}
respectively.

We note that the model with only the predation term reduces to the jump-growth equation derived in \cite{Datta2008} when the feeding rate is chosen as
$\wp(w,w',w'')=k(w,w')\delta(w+Kw'-w'')$, for some feeding preference function $k(w,w')$ and a fraction $K$ accounting for the feeding efficiency.
As pointed out in \cite{Datta2008}, when the typical predator:prey mass ratio is sufficiently large and the population density is sufficiently smooth that model in turn can be approximated by the McKendrick-von Foerster equation \cite{McKendrick1925,vonFoerster1959}. It is this simplified model in terms of the McKendrick-von Foerster equation that forms the basis of most
previous analytic studies of the size distribution \cite{SilvertPlatt1978,SilvertPlatt1980,Cushing1992,CamachoSole2001,BenoitRochet2004,Arino2004}. We will see that, thanks to scale invariance, the much
more general evolution equation \eqref{eq:detcon} is amenable to similar analytical investigations.

\section{\label{s:scale}Scale invariance}

We now derive restrictions on the feeding rate $\wp$, the reproductive rate $\wt$, the maintenance rate $\wa$ and the mortality rate $\wb$ that guarantee the scale invariance of our model.

As we have a model describing the dependence of the abundance on two variables, namely weight and time, we can consider separate scaling in weight and in time. However, we expect our model to be invariant only under simultaneous scaling of both weight and time, because we expect life processes to run faster for smaller organisms. Hence, when we change the weight
scale by a factor $\va$ we should also change the time scale by a factor $\va^\xi$, where the constant $\xi$ expresses how the speed of the dynamics scales with weight. So we will
consider the transformations
\begin{equation}\label{eq:cxi}
(w,t) \mapsto (\va w, \va^\xi t) \text{ with } \va > 0.
\end{equation}
Under such scale transformations the density $\phi(w,t)$ transforms as
\begin{equation}\label{eq:phit}
\phi(w,t)\mapsto \va^\gamma\phi(\va w, \va^\xi t)
\end{equation}
with some, so far undetermined, exponent $\gamma$, also called the scaling dimension of $\phi$ \cite{DiFrancesco1997}.

Requiring the evolution equation \eqref{eq:detcon} to remain unchanged under this transformation imposes the conditions
\begin{equation}
\begin{split}\label{eq:rtx}
 \va^{\gamma-\xi-2}\wp(w,w',w'') &=  \wp(cw,cw',cw''),\\
 \va^{-\xi-2}\wt(w,w',w'') &=  \wt(cw,cw',cw''),\\
 \va^{1-\xi} \wa(w) &= \wa(cw),\\
 \va^{-\xi} \wb(w) &=\wb(cw).
\end{split}
\end{equation}

We choose to factorise the feeding rate as
\begin{equation}\label{eq:pfact2}
 \wp(w,w',w'') = \ws(w,w')\wg(w,w',w''),
\end{equation}
where $\ws(w,w')$ determines the rate at which a predator of weigh $w$ eats a prey of weight $w'$ and $\wg(w,w',w'')$ gives the probability density that such a feeding event makes the predator grow to weight $w''$.
On average only a certain proportion $Q$ of the prey biomass will be absorbed by the predator, so that on average $w''=w+Qw'$, but there will be some variability, due to differences in predator digestion and prey composition. This will be modelled by the probability density $\wg(w,w',w'')$, so that
\begin{equation}\label{eq:q}
 \int dw''\ w'' A(w,w',w'') = w+Qw'.
\end{equation}
We should make it clear that the model, and therefore the results, depend only on the combined rate $\wp$. We factorise it into $\ws$ and $\wg$ only to make the ecological origin of the rate clearer \cite{BenoitRochet2004,Datta2008}, but how we choose this factorisation has no influence on the results.

Because the model only depends on the product of $\ws$ and $\wg$, we are free to choose the relative scaling between these factors, as long as the product scales as in \eqref{eq:rtx}. We want $\wg$ to scale as a probability density in $w''$ and hence we choose
\begin{equation}\label{eq:gsci}
\begin{split}
 \wg(w,w',w'')&= w''^{-1}\sg(w/w',w''/w),\\
 \ws(w,w') &= \left({w}/{w_0}\right)^{\gamma-\xi-1}\ss(w/w'),
\end{split}
\end{equation}
Here $w_0$ is an arbitrarily chosen reference weight.
We have also introduced the scaling
functions $\sg$ and $\ss$ that are invariant under scale transformations.

Similarly, we require that the reproduction rate $\wt$ scale as a density in $w'$ and $w''$, so according to the scaling rules given in \eqref{eq:rtx} we can write
\begin{equation}\label{eq:qsci}
 \wt(w,w',w'') = w'^{-1}w''^{-1}\left({w}/{w_0}\right)^{-\xi}\st(w/w',w''/w)
\end{equation}
for some scaling function $\st$. This behaviour for the reproduction rate implies that the average weight of an offspring is proportional to $w^{1-\xi}$, where $w$ is the
weight of its parent. This should however not be taken as a prediction of the scaling of egg sizes because, as stressed previously, our model of reproduction is not formulated at a sufficient level of detail for that purpose.

The scale transformations for the maintenance and death rates given in \eqref{eq:rtx} allow us to express them as
\begin{equation}\label{eq:sip}
\begin{split}
 \wa(w)&=\left({w}/{w_0}\right)^{1-\xi}\wa(w_0),\\
 \wb(w) &= \left({w}/{w_0}\right)^{-\xi}\wb(w_0),
\end{split}
\end{equation}

The functions $\ss$, $\sg$ and $\st$, the constants $\wa(w_0)$ and $\wb(w_0)$ and the exponents $\gamma$ and $\xi$ are not fixed by the requirement of scale invariance and need to be
determined from observations or separate theoretical arguments.

The restrictions that we impose to achieve scale invariance predict the relative scaling of
these rates. In particular, the scaling of all the rates contains the exponent $\xi$.
It is widely believed that the maintenance metabolic rate scales as $w^{3/4}$
\cite{ClarkeJohnston1999,Brown2004,Peters1986}. Comparing that to the scaling in \eqref{eq:sip} gives $\xi=1/4$. This then predicts that the mortality rate scales as $w^{-1/4}$, which is in line with observations \cite{Peters1986,Lorenzen1996}. Moreover, the
exponent $\gamma$ of the scaling transformation \eqref{eq:phit} of the density is determined once the scaling of the
feeding rate and $\xi$ are known. If we assume, following Beno\^{\i}t and Rochet \cite{BenoitRochet2004},
that the feeding rate
is proportional to the volume searched per unit time, we can identify the scaling exponent of $S$ as the exponent
of the search volume, which is around 0.8 \cite{Ware1978}. Assuming $\xi=1/4$ it follows that $\gamma\approx 2$, which is in line with observations as well \cite{Blanchard2009}. We will see in Section~\ref{s:ss} that the exponent
$\gamma$ coincides with the exponent of the power-law steady-state solution. Therefore, the relative scaling between density-dependent processes (feeding) and density-independent processes (reproduction, maintenance and death) determines the steady state exponent.

Besides scale invariance, our model also has time translation invariance. Time translations and scale transformations together generate the non-abelian symmetry group
${\rm Aff}(1,\mathbb{R})$, the group of orientation-preserving linear transformations of the real line, also known as the $ax+b$ group. In the special case $\xi=0$ the symmetry group
becomes abelian, a point which we will exploit in Section~\ref{s:stability}, because in that case the equation is
translationally invariant and hence the standard Fourier transform can be applied to solve the equation for small perturbations of the steady-state.

The symmetry is enhanced when the model only contains the predation terms, which are quadratic in $\phi$. In that case there is also an invariance under scaling in time alone,
\begin{equation}
 \phi(w,t)\mapsto\lambda\phi(w,\lambda t)
\end{equation}
for any $\lambda\in\mathbb{R}$.

\section{\label{s:log}Change to logarithmic weight}

For the upcoming analysis it is convenient to make a change of variable to $x = \log(w/w_0)$, where $w_0$ is the arbitrarily chosen reference weight. We refer to $x$ as the logweight.
A scale transformation $w\mapsto c w$ corresponds to a translation $x\mapsto x+\log c$.

We introduce the density $u(x)$ so that $\V u(x)dx$ is the number of individuals in volume $\V$ with a logweight between $x$ and $x+dx$. Thus $u(x)=w\phi(w)$ and we can easily
translate our results for $u(x)$ back into results for
$\phi(w)$, if desired. Under a scale transformation the function $u(x,t)$ transforms to $c^{\gamma-1} u(x+\log c, c^\xi t)$. We will apply this change to the various terms of the evolution equation \eqref{eq:detcon}.

We first introduce the predation rate $\xp$ such that $\xp(x,x',x'')=w''\wp(w,w',w'')$. The factorisation of $\wp$ into $\ws$ and $\wg$ in equation \eqref{eq:pfact2}
together with equation \eqref{eq:gsci} leads to
\begin{equation}\label{eq:psa}
\xp(x,x',x'')=e^{(\rho-\xi)x}\xs(x-x')\xg(x-x',x''-x),
\end{equation}
where we have defined the functions $\xs(y)=\ss(e^y)$ and $\xg(y,z)=\sg(e^y,e^{z})$ and we have introduced the exponent $\rho=\gamma-1$ for latter convenience. We now transform to logweights, substitute this form for $\xp$ into \eqref{eq:detconpred} and
perform a change of variables in each of the feeding terms so that the integration variable coincides with the argument of $\xs$. The result is
\begin{widetext}
\begin{equation}\label{eq:logtransp}
\left(\frac{\partial u(x)}{\partial t}\right)_{\rm\wp}
=e^{(\rho-\xi)x}\int dy\ \xs(y)\left(-u(x)u(x-y)-e^{(\rho-\xi)y}u(x)u(x+y)
+\int dz\ e^{-(\rho-\xi)z}\xg(y,z)u(x-z)u(x-y-z)\right),
\end{equation}
where we have taken into account the fact that $\xg(y,z)$ is a probability density and hence it is normalised. The integrals all run over the whole real line. In what follows,
we will often not indicate the time-dependence of $u(x,t)$ explicitly but write just $u(x)$ instead, as in the above equation.

We can perform the same changes for the reproduction term. Using the scaling form \eqref{eq:qsci} for $\wt$ in \eqref{eq:detconrep}, then transforming to logweights and defining
$\xt(y,z)=\st(e^y,e^z)$, we obtain
\begin{equation}\label{eq:logtransr}
\left(\frac{\partial u(x)}{\partial t}\right)_{\rm\wt}=
e^{-\xi x}\int dy\int dz\ \xt(y,z)\Big(-u(x)+e^{-\xi y}u(x+y)
+e^{(\xi-1)z}(1-e^{-y})u(x-z)\Big).
\end{equation}
\end{widetext}

Finally, the contribution of maintenance and intrinsic mortality can be expressed in logarithmic weights as
\begin{equation}\label{eq:logtransma}
\left(\frac{\partial u(x)}{\partial t}\right)_{\rm\wa}=
\xa\frac{\partial}{\partial x}\left[e^{-\xi x}u(x)\right]
\end{equation}
and
\begin{equation}\label{eq:logtransmo}
\left(\frac{\partial u(x)}{\partial t}\right)_{\rm\wb}=-\xb e^{-\xi x} u(x),
\end{equation}
respectively. Here we have introduced the constants $\xa=\wa(w_0)/w_0$ and $\xb=\wb(w_0)$.

The full dynamical equation in terms of logweights is
\begin{equation}\label{eq:logtrans}
\frac{\partial u}{\partial t}=\left(\frac{\partial u}{\partial t}\right)_{\rm\wp}
+\left(\frac{\partial u}{\partial t}\right)_{\rm\wt}+\left(\frac{\partial u}{\partial t}\right)_{\rm\wa}
+\left(\frac{\partial u}{\partial t}\right)_{\rm\wb}.
\end{equation}
We have chosen not to fully non-dimensionalise this evolution equation: $t$ still has dimension of time, $u(x)$ has the dimension of inverse volume, $\xa$, $\xb$ and $\xt$ have
dimension of inverse time, $\xs$ has dimension of volume over time and $\xg$ is dimensionless.

\section{\label{s:ss}Power-law steady state solution}

Solving the integro-differential evolution equation \eqref{eq:logtrans} is difficult in general. However we can simplify the task by looking for solutions that are
invariant under symmetry transformations.

In this section we will study the solution that is invariant under both time-translations and scale transformations. We leave the study of the general scale-invariant solutions to Appendix~\ref{s:sst}. Invariance under time-translations means that we are looking
for a steady-state solution $\hat{\phi}(x)$ that has no dependence on time. Invariance under scale transformations \eqref{eq:cxi} and \eqref{eq:phit} then implies
\begin{equation}\label{eq:ssw}
  \hat{\phi}(w)=\hat{\phi}(w_0) \left(\frac{w}{w_0}\right)^{-\gamma},
\end{equation}
where $w_0$ is the arbitrarily chosen reference weight.
After transforming to logweights as in Section~\ref{s:log} this takes the form
\begin{equation}\label{eq:ssx}
\hat{u}(x)=u_0e^{-\rho x},
\end{equation}
where $u_0=w_0 \hat{\phi}(w_0)$ and $\rho=\gamma-1$. Substituting this form for the solution into the evolution equation \eqref{eq:logtrans} gives an equation for the overall population level $u_0$,
\begin{equation}\label{eq:stecon1}
 c_{\rm\wp}u_0=-\xa(\rho+\xi)-\xb+c_{\rm\wt},
\end{equation}
where the constants $c_{\rm\wp}$ and $c_{\rm\wt}$ are given by
\begin{equation}\label{eq:cc}
\begin{split}
 c_{\rm\wp}&=\int dy\ \xs(y)\left(e^{\rho y}+e^{-\xi y}
 -e^{\rho y}\int dz\ e^{(\rho+\xi)z}\xg(y,z)\right)\\
 c_{\rm\wt}&=\int dy\int dz\ \xt(y,z)\Bigl(-1+e^{-(\rho+\xi) y}\\ &\hspace{3.5cm}+e^{(\rho+\xi-1)z}(1-e^{-y})\Bigr).
\end{split}
\end{equation}
When $c_{\rm\wp}\neq 0$ this uniquely determines $u_0$ and hence the steady state solution.
We will show in Appendix~\ref{s:uzero} that $u_0$ is positive under some ecologically
reasonable assumptions about the parameter functions.

Note that scale invariance fixes the power-law form of the steady-state size spectrum and the steady-state exponent $\rho$ is determined entirely by the scaling behaviour of the parameter functions, see \eqref{eq:rtx}.
It is not dependent on any other details of the interactions in the model.

A special situation arises in the case where maintenance, reproduction and intrinsic mortality are absent from the model. In this case only the first scaling relation in
\eqref{eq:rtx} remains and it is not enough to determine both $\xi$ and $\rho$. However an equation for $\rho$ is obtained by noticing that in this case the right-hand side of
\eqref{eq:stecon1} is zero and for $u_0\neq 0$ this implies that $c_{\rm\wp}=0$.
This constraint should then be used to determine the scaling exponent $\rho$ given a particular choice for $s(y)$ and $a(y,z)$. The overall population level $u_0$ is not
determined by the model in this case. This special situation was considered in most previous work \cite{SilvertPlatt1978,SilvertPlatt1980,BenoitRochet2004,Datta2008}.

\subsection{Conservation of number of individuals}

There is a continual flux of individuals from lower weight to
larger weight to make up for the losses due to predation and intrinsic death.
In previous models that considered only the predation process \cite{BenoitRochet2004,Datta2008} there was no source for this influx of small individuals. Instead they appeared from $x=-\infty$. Now that we are modelling the reproduction process, we do have a source of individuals and can impose that in the steady state this source should exactly balance the losses.

The easiest way to impose this balance is to impose for each weight bracket $i$ that the number of individuals entering the bracket from the left due to predation exactly equals
the number of individuals leaving that bracket to the left, either as offspring or through weight-loss. This gives
\begin{equation}
 \sum_{j,k}\wp_{jki}\hat{\phi}_j\hat{\phi}_k=\sum_{j,k}(m_{ijk}+1)\wt_{ijk}\hat{\phi}_i
 +\wa_i\hat{\phi}_i.
\end{equation}
Thanks to scale invariance, all these conditions for different $i$ are equivalent.
In the continuum, after substituting the steady state solution and changing to logweight notation, this condition reads
\begin{equation}\label{eq:flux}
f_{\rm\wp}u_0=f_{\rm\wt}+\xa,
\end{equation}
where we have defined the constants
\begin{equation}\label{eq:ff}
\begin{split}
f_{\rm\wp} &=\int\! dy\!\int\! dz\, \xs(y)\,\xg(y,z)e^{(\rho+\xi)z+\rho y},\\
f_{\rm\wt} &=\int\! dy\!\int\! dz\, \xt(y,z) \left(e^{-z}(1-e^{-y})+1\right).
\end{split}
\end{equation}
We can use this constraint, together with the steady-state condition \eqref{eq:stecon1} to fix the maintenance rate in terms of the other parameters of the model,
\begin{equation}\label{eq:r}
\xa = \frac{(c_{\rm\wt}-d)f_{\rm\wp}-c_{\rm\wp}f_{\rm\wt}}{c_{\rm\wp}+(\rho+\xi)f_{\rm\wp}}.
\end{equation}
If we use this to eliminate $\xa$ from the equation \eqref{eq:stecon1} for $u_0$ we get
\begin{equation}\label{eq:u0}
u_0 = \frac{c_{\rm\wt}-d+(\rho+\xi)f_{\rm\wt}}{c_{\rm\wp}+(\rho+\xi)f_{\rm\wp}}.
\end{equation}

Obviously, for the model to make sense we need $\xa$ to be positive. This is not possible for all choices of the other parameters. In particular, this requirement defines a range of allowed exponents $\rho$. To investigate this further, we will now make specific choices for the functions that appear in the feeding and reproduction rates.

\subsection{\label{ss:param}Choice of parameter functions}

For the prey selection function $\xs(y)$ we choose a Gaussian that expresses that there is a preferred value $\beta$ for the log of the predator:prey mass ratio and a certain
variance $\sigma_\beta^2$ around this mean \cite{Ursin1971}. So we set
\begin{equation}\label{eq:gaufeed}
\xs(y)=s_0 g_{\sigma_\beta}(y-\beta)
\end{equation}
with
\begin{equation}\label{eq:gauker}
 g_\sigma(x)=\frac{1}{\sqrt{2\pi}\sigma}e^{-x^2/2\sigma^2}.
\end{equation}
The parameter $s_0$ has dimension of volume over time and sets the overall feeding rate.

For the absorption probability density $\xg$ the simplest assumption would be that a fixed proportion $Q$ of prey mass is absorbed in all feeding events, i.e., that in terms of the
predator mass $w$ and the prey mass $w'$ the mass after feeding is always $w''=w+Qw'$. This corresponds to a choice $\xg(y,z)=\delta(z-\psi(y))$
where
\begin{equation}
 \psi(y)=\log(1+Qe^{-y}).
\end{equation}
This was used in \cite{Datta2008}.
However the proportion of the prey mass that is absorbed by the predator is not exactly the same in each feeding event. Variability arises for example from the difference in
digestion between predator species and also from the difference in organic composition of prey organisms. In this paper we will allow variation by replacing the delta function by
a Gaussian. So we will set
\begin{equation}\label{eq:genker}
 \xg(y,z) = g_{\sigma_\psi}(z-\psi(y)).
\end{equation}

For the reproduction function $\xt(y,z)$ we will use the product of Gaussians
\begin{equation}\label{eq:t}
  \xt(y,z)=\xt_0 g_{\sigma_\nu}(y-\nu) g_{\sigma_\mu}(z-\mu).
\end{equation}
This gives a mean offspring:parent mass ratio of $e^\mu$ and a mean mass ratio between parent before reproduction and parent after reproduction of $e^\nu$.

\begin{figure}
 \includegraphics[width=6cm]{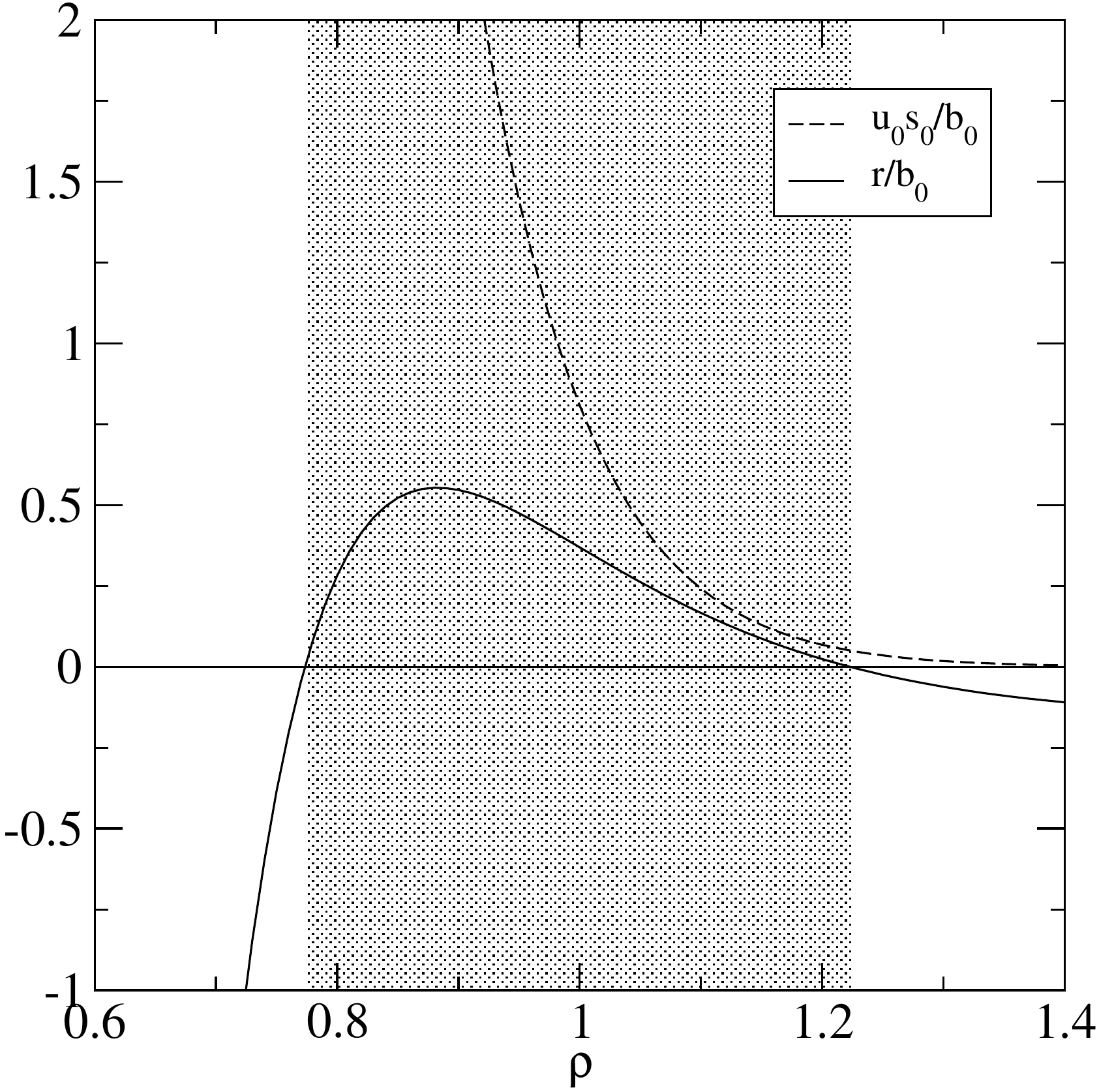}
 \caption{\label{fig:coef3}Plot of the curves for $u_0$ and $\xa$ given in \eqref{eq:u0} and \eqref{eq:r} with the choices \eqref{eq:gaufeed}, \eqref{eq:genker} and \eqref{eq:t},
for $\xi=0.25$. The rest of the parameters are $\beta=5.5$, $\sigma_{\beta}=2.5$, $\sigma_{\psi}=0$, $\mu=-10$, $\sigma_{\mu}=0.5$, $\nu=0.2$,
$\sigma_{\nu}=0.05$ and $d=0$. The region of allowed $\rho$ is shaded.}
\end{figure}

\begin{figure*}
 \includegraphics[width=12cm]{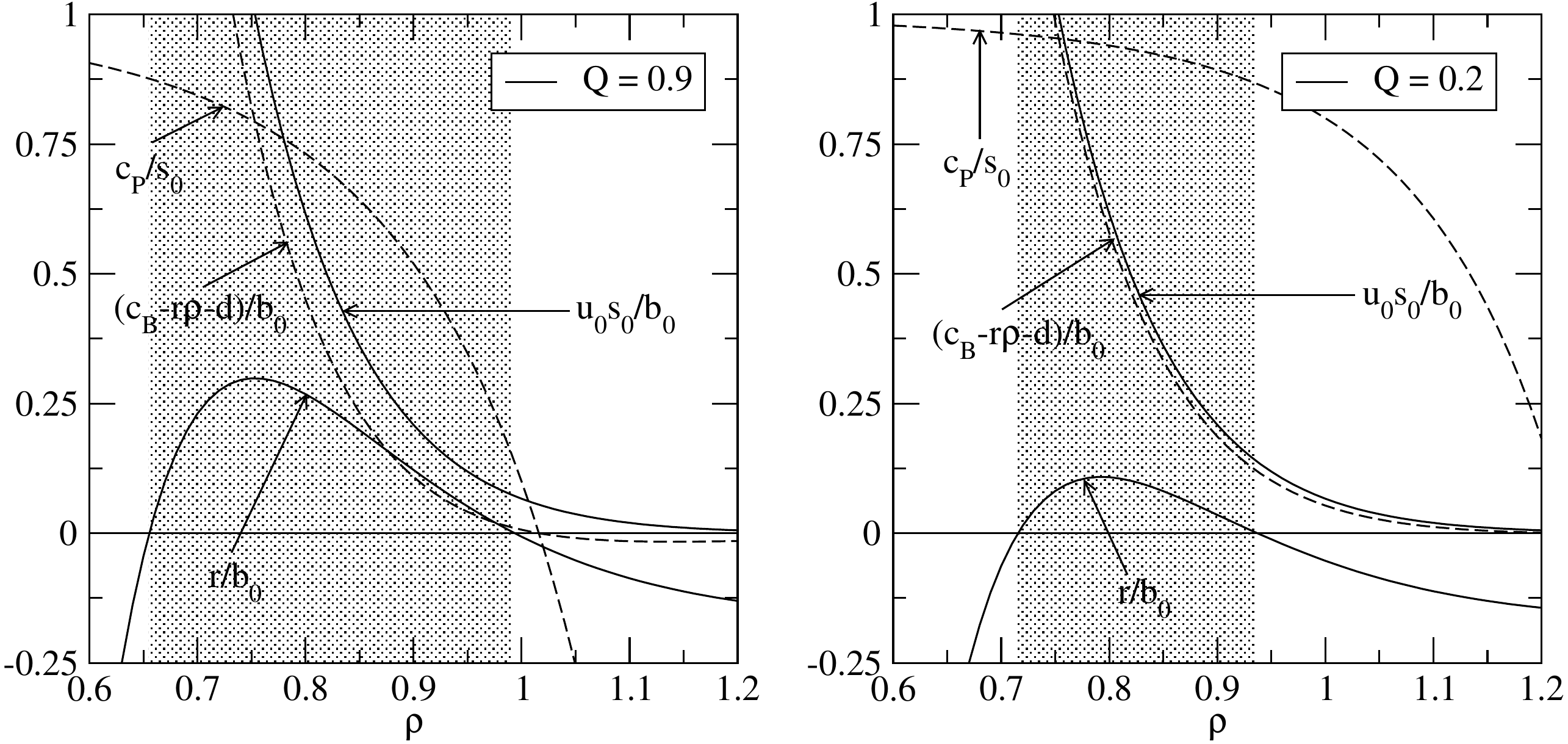}
 \caption{\label{fig:coef}Same as Figure~\ref{fig:coef3} but with $\xi=0$ and
for two different values of $Q$. All the remaining parameters remain unchanged except for $\mu=-7.5$. Shaded regions show the intervals where $\xa$ is positive,
which determines regions of allowed $\rho$. We find that both $c_{\rm\wp}$ and
$c_{\rm\wt}-\xa\rho-\xb$ are positive as well within that interval, leading to stable values of $u_0$.}
\end{figure*}

Finding the correct values for the parameters requires a close investigation of the data and is outside the scope of this paper. However, for the purpose of the plots, we have
chosen parameters that appear at least reasonable from a biological point of view. For example, the preferred predator:prey body mass ratio is believed to be around $10^2$ or $10^3$
\cite{JenningsMackinson2003}, so we have chosen $\beta=5.5$. In order to estimate $\nu$, we need the average weight loss
caused by reproduction processes.
The average gonadosomatic index (ratio between the gonadal weight and the body weight) is actually measured for fish
and is rather variable. It is found to be on average around 0.1 or 0.2 \cite{Roff1992},
thus we have chosen $\nu=0.2$ so that the average fraction of weight loss due to reproduction is around 20\%.
We have set the logweight difference between offspring and parent $\mu$ to a small value around -8 or -9. For the standard deviations in the parameters we use
$\sigma_{\beta}=2.5$, $\sigma_{\nu}=0.05$ and $\sigma_{\mu}=0.5$, although a careful analysis of the data will be necessary to determine them properly.
We have set the absorption efficiency to $Q=0.9$ \cite{PandianMarian1985} because respiration and other metabolic processes
have been modelled separately. In previous work \cite{Datta2010}, where these processes were not separated, the net absorption efficiency was replaced by a conversion efficiency
of around $0.2$. In most plots we will set the variability in $Q$ equal to zero, as well as the mortality rate.

In Figure~\ref{fig:coef3} we have plotted the maintenance rate $\xa$ and the steady-state density coefficient $u_0$ as functions of $\rho$ for the above choices of the parameters and $\xi=0.25$. The allowed
interval for $\rho$ appears shaded in that figure. It is encouraging that the observed value $\rho\approx 1$ \cite{Sheldon1972,KerrDickie2001,JenningsMackinson2003} is contained within the interval.

\section{\label{s:stability}Stability of the steady state}

It has been observed via numerical simulations in \cite{Law2009,Datta2008} that the power-law steady state is not always stable against small perturbations but rather that the
system can undergo a bifurcation in which the steady state becomes unstable and a stable travelling wave solution emerges. This phenomenon was investigated analytically in
\cite{Datta2010} through a linear stability analysis. We now perform a similar analysis in our generalised model.

The only other paper that we are aware of that investigates the stability of the power-law steady state is \cite{Arino2004} but it deals, for reasons of simplicity, with a model
where the growth due to feeding is independent of the prey density, thus avoiding having to deal with the associated non-linear terms.

In order to discuss stability analytically, we consider the particular case $\xi=0$. According to \eqref{eq:sip}, this corresponds to a maintenance rate proportional to the weight
and a mortality rate independent of the weight. This is not quite realistic, but simplifies the analysis considerably because it leads to translational invariance in both time and
logweight. This will allow us to use the standard Fourier transform.

\subsection{Perturbation in $u_0$}

Before we consider the general, weight-dependent perturbation we take a look at a particular perturbation that affects only the overall population density $u_0$. So instead of
\eqref{eq:ssx} we consider the solution
\begin{equation}
 u(x,t)=u_0(t)e^{-\rho x}
\end{equation}
where we now allow $u_0$ to depend on time. Substituting this into the evolution equation \eqref{eq:logtrans} with $\xi=0$ gives
\begin{equation}\label{du0}
 \frac{du_0}{dt}=-c_{\rm\wp}u_0^2+(c_{\rm\wt}-r\rho-d)u_0,
\end{equation}
where $c_{\rm\wp}$ and $c_{\rm\wt}$ are given in eq. \eqref{eq:cc}. The solutions to this differential equation depend crucially on the strength
of reproduction relative to maintenance and mortality. If reproduction is weak, i.e., if $c_{\rm\wt}<r\rho+d$, then the non-zero fixed point (steady state) at
$u_0=(c_{\rm\wt}-r\rho-d)/c_{\rm\wp}$ is unstable. If, however, reproduction is strong enough so that
\begin{equation}\label{eq:us}
c_{\rm\wt}>r\rho+d,
\end{equation}
then the fixed point is stable. In between, the system undergoes a bifurcation at which there is a whole line of fixed points exactly when $c_{\rm\wt}=r\rho+d$.

In particular, this observation shows that the model without reproduction can never be stable against a perturbation in the overall population density $u_0$. This instability was
already noticed in \cite{Datta2010}, where it was argued that it could be avoided by suitable modifications of the model at the ends of the size-spectrum (plankton dynamics,
senescent death). The observation that reproduction can provide a stabilising effect is new to this paper.

As we have seen, the stability against variations in $u_0$ requires that $c_{\rm\wp}$ and $c_{\rm\wt}-\xa\rho-\xb$ are positive. In Figure~\ref{fig:coef} we plot these coefficients together with $u_0$ and $\xa$ as functions of $\rho$, for two different
values of $Q$. As discussed in Section~\ref{s:ss}, the parameter $\rho$ is only allowed to lie in a certain interval where the maintenance rate is positive, which appears shaded.
Within this interval, both $c_{\rm\wp}$ and
$c_{\rm\wt}-\xa\rho-\xb$ are seen to be positive, so steady state is stable in this case.

\begin{figure}
 \includegraphics[width=6cm]{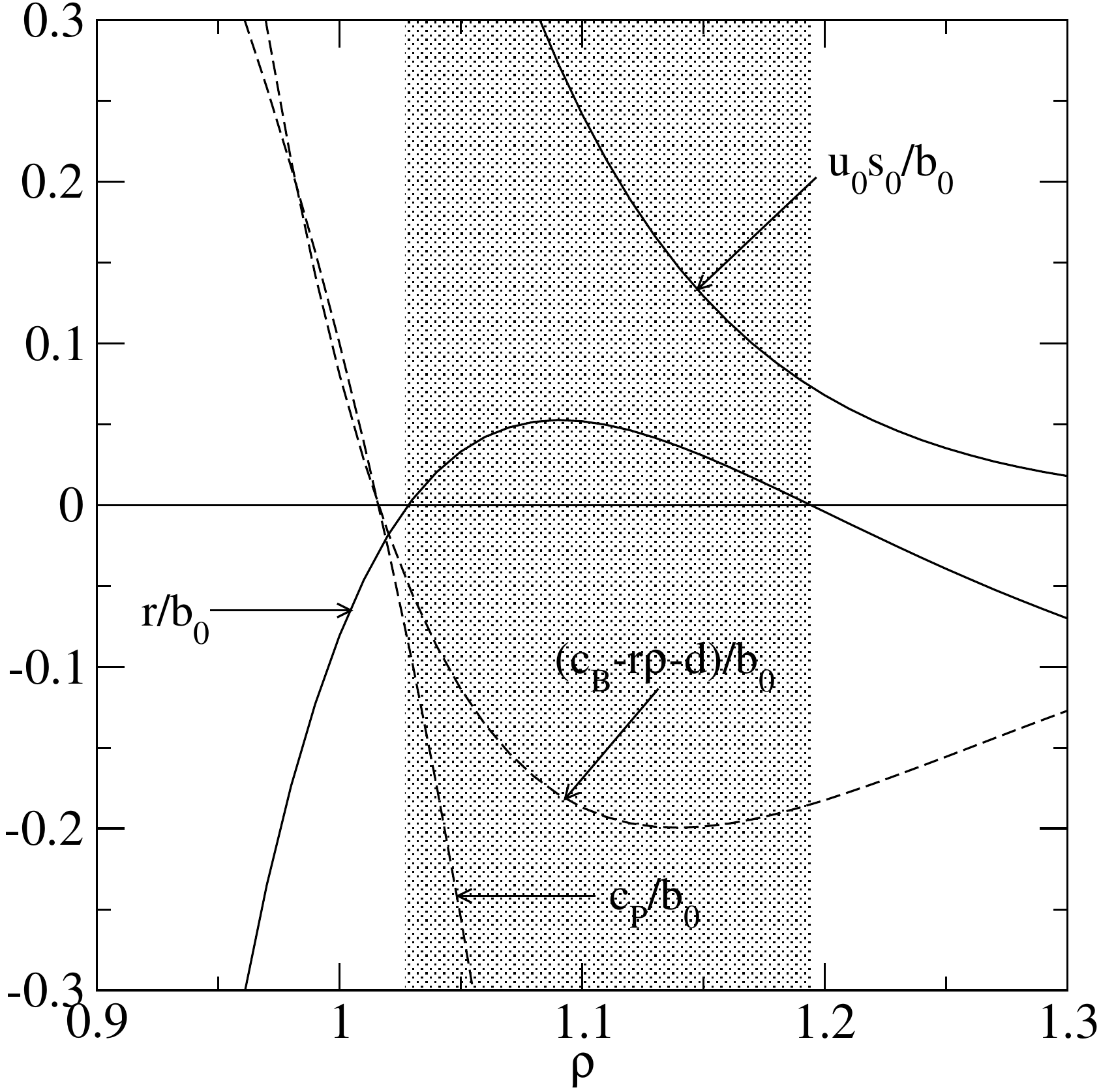}
 \caption{\label{fig:coef2}Same as Figure~\ref{fig:coef} left, except that $\mu=-10$. Shaded region show the interval for allowed $\rho$ but unstable $u_0$ (note that
both $c_{\rm\wp}$ and $c_{\rm\wt}-\xa\rho-\xb$ are negative within the interval).}
\end{figure}

The region where both $c_{\rm\wp}$ and $c_{\rm\wt}-\xa\rho-\xb$ are negative corresponds to an unstable steady state, as shown in Figure~\ref{fig:coef2} with the same parameters
but $\mu=-10$. For the following plots we will choose $\rho=0.9$ as a suitable value leading to a stable steady state (with $\mu\approx -8$) and for the unstable solution we will
choose $\rho=1.1$ (for $\mu\approx -10$).

\subsection{General perturbation}

\begin{figure}
 \includegraphics[width=6.5cm]{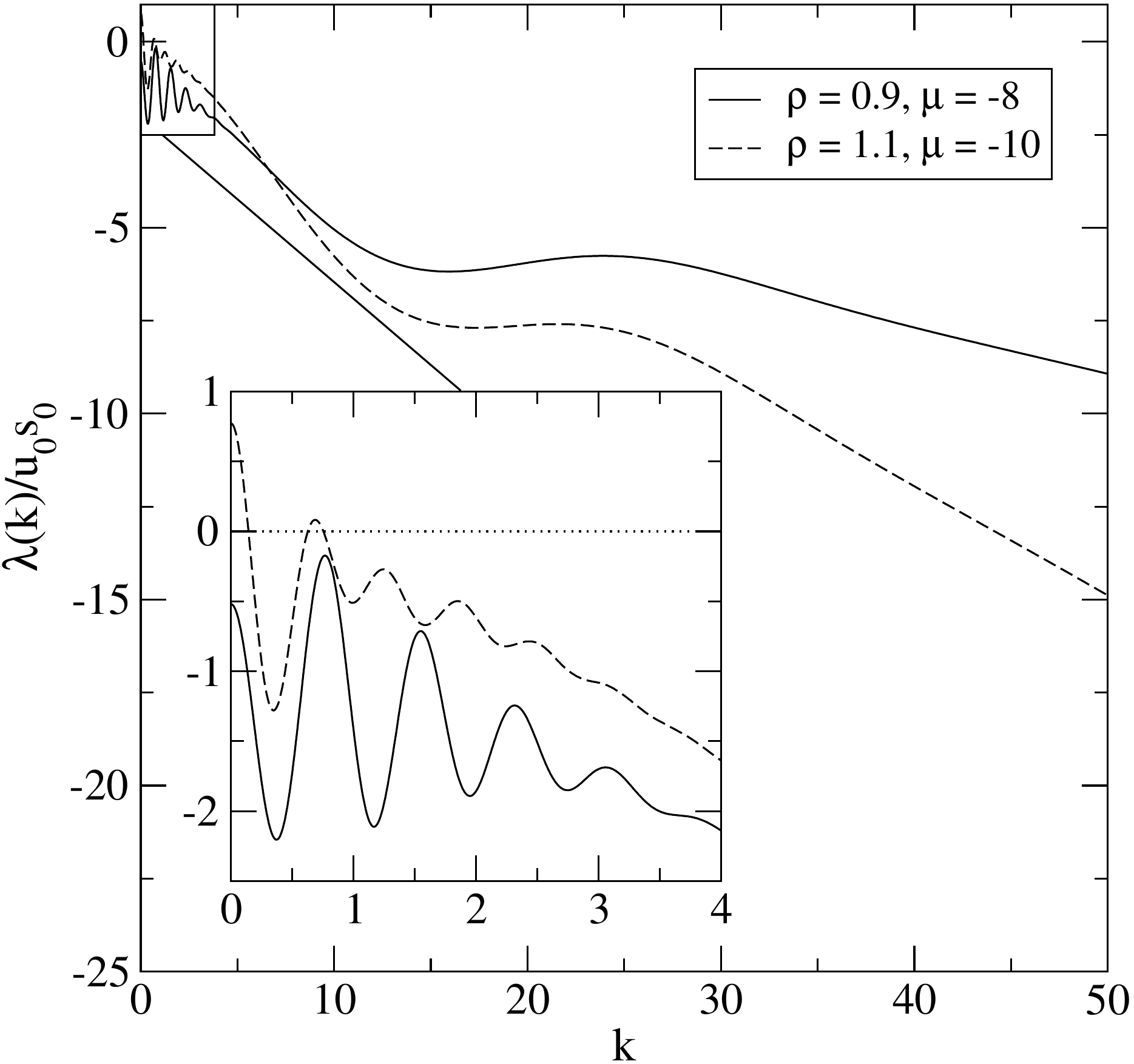}
 \caption{\label{fig:lamPB}The combined spectrum $\lambda(k)$. Remaining parameters are: $\beta=5.5$, $\sigma_{\beta}=2.5$, $Q=0.9$, $\sigma_{\psi}=0$,
$\sigma_{\mu}=0.5$, $\nu=0.2$ and $\sigma_{\nu}=0.05$. Note that $\rho=0.9$ lies in the region of stable $u_0$, whereas the fixed point for $u_0$ is unstable when $\rho=1.1$.
Inset contains the region of small $k$.}
\end{figure}

\begin{figure*}
 \includegraphics[width=14cm]{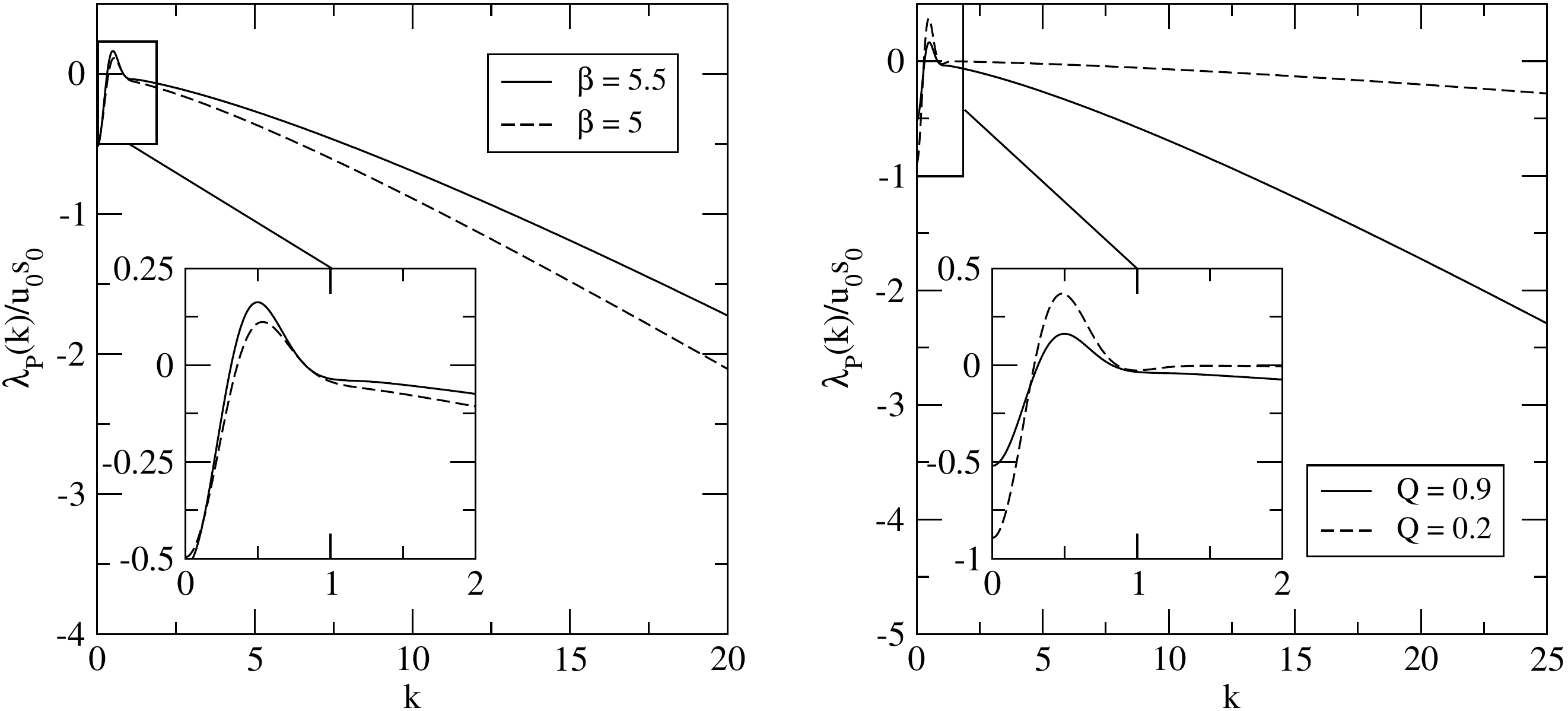}
 \caption{\label{fig:lamPbetaQ}Eigenvalue spectra $\lambda_{\rm\wp}$ for two different values of $\beta$ with $Q=0.9$ (left), and for two different values of $Q$ with $\beta=5.5$ (right).
Remaining parameters are: $\sigma_{\beta}=2.5$, $\sigma_{\psi}=0$ and $\rho=0.9$.  Insets contain zooms of the regions with small $k$.}
\end{figure*}

\begin{figure}
 \includegraphics[width=6.5cm]{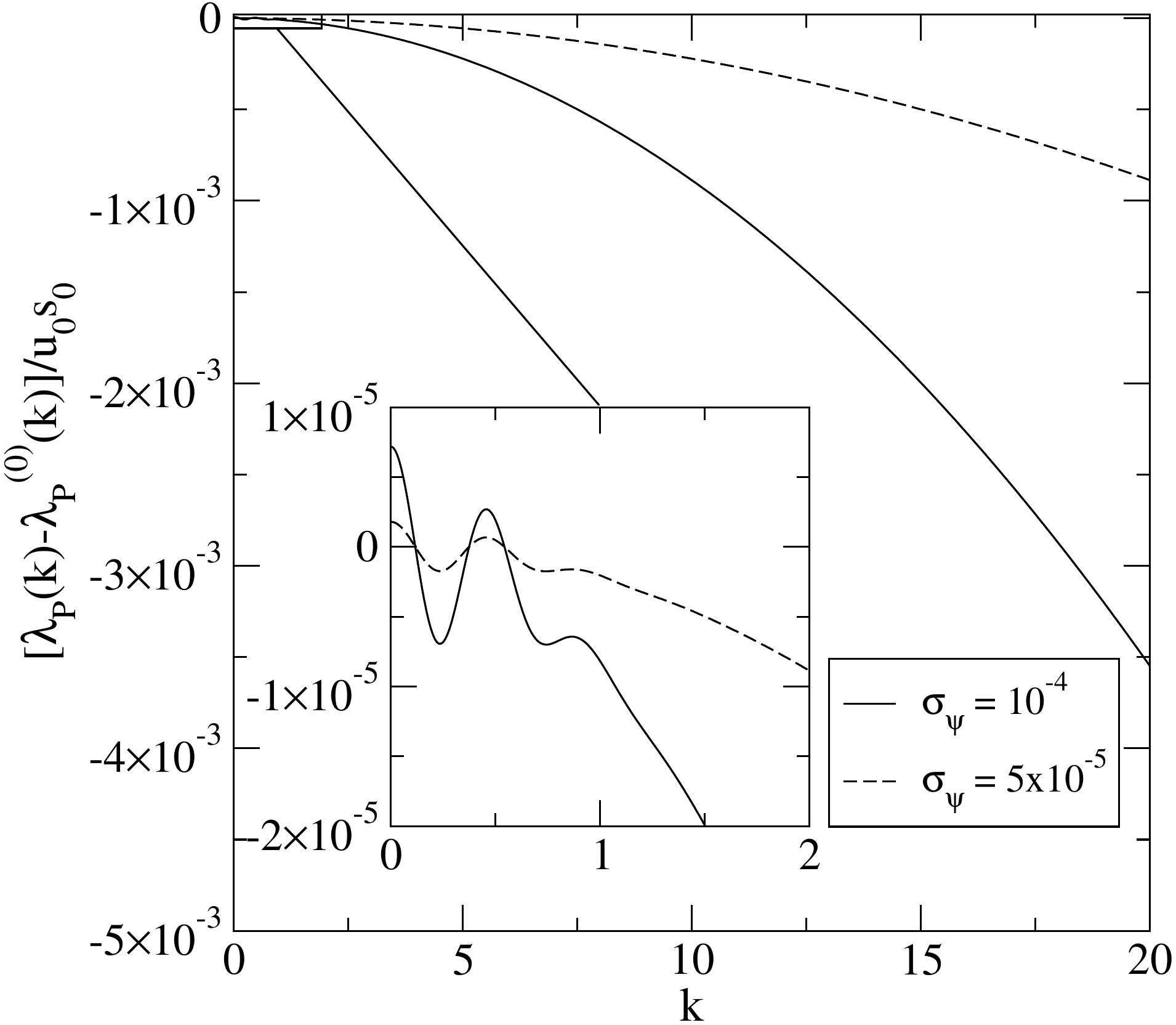}
 \caption{\label{fig:lamspsi}Effect of the variability in $Q$. Here we plot the difference $\lambda_{\rm\wp}(k)-\lambda_{\rm\wp}^{(0)}(k)$, where $\lambda_{\rm\wp}^{(0)}(k)$ is the
value for $\sigma_{\psi}=0$. In the inset we plot a zoom of the spectrum for low wavenumbers. Note that the magnitude of the effect of the variability in $Q$ is very small. Remaining
parameters are: $\beta=5.5$, $\sigma_{\beta}=2.5$, $Q=0.9$ and $\rho=0.9$. Inset contains the zone of small wavenumbers.}
\end{figure}

\begin{figure*}
 \includegraphics[width=14cm]{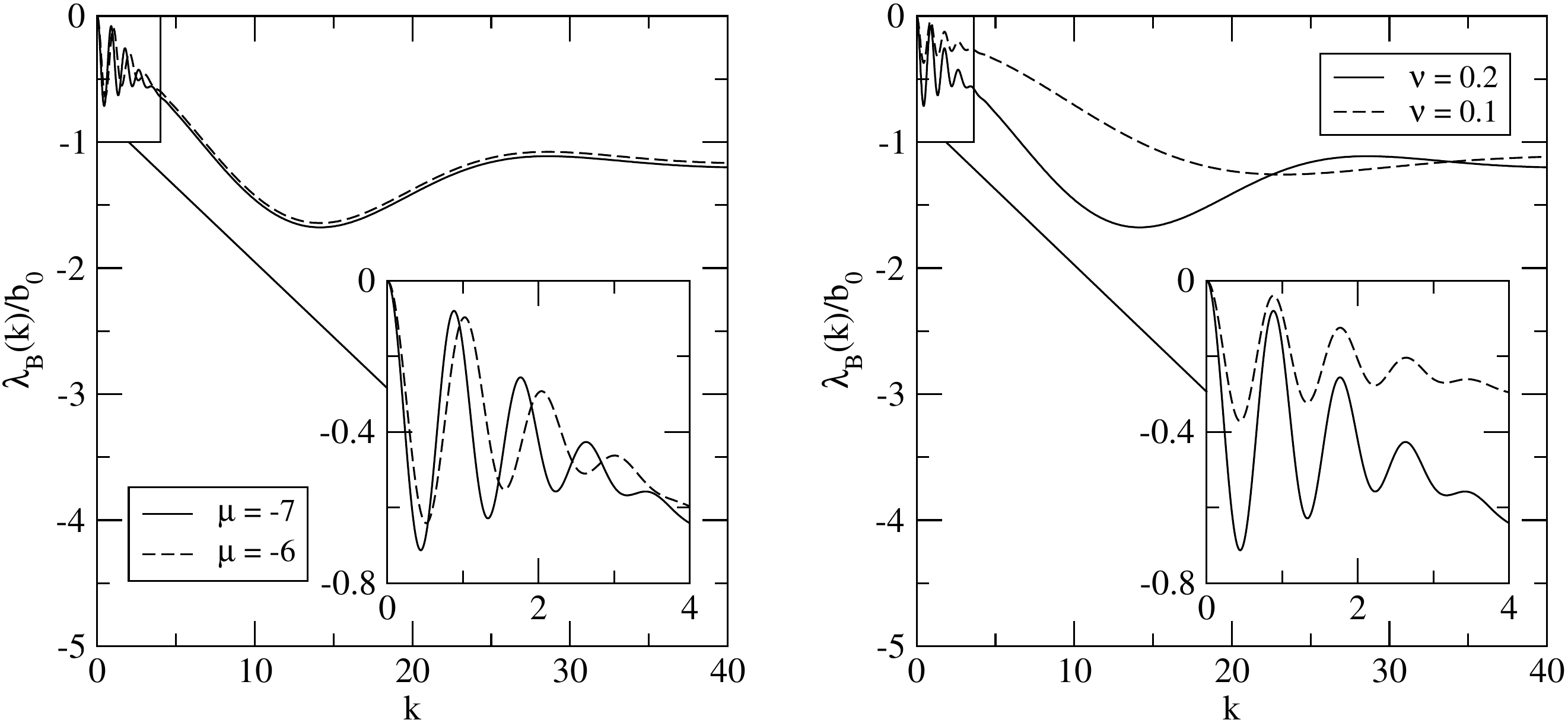}
 \caption{\label{fig:lamBmunu}Eigenvalue spectra $\lambda_{\rm\wt}$ for two different values of $\mu$ with $\nu=0.2$ (left), and for two different values of $\nu$ with $\mu=-7$
(right). Remaining parameters are: $\sigma_{\mu}=0.5$, $\sigma_{\nu}=0.05$ and $\rho=0.9$. Insets contain zooms of the regions with small $k$.}
\end{figure*}

We now consider more general, weight-dependent perturbations.
It is convenient to change to the new density $v$ related to $u$ through \begin{equation}
 u(x,t)=u_0e^{-\rho x}v(x,t),
\end{equation}
so that the steady-state solution is just a constant $v(x,t)=1$. We add a small perturbation $\epsilon(x,t)$ to the steady-state solution
\begin{equation}
v(x,t)=1+\epsilon(x,t)
\end{equation}
and linearise the equation \eqref{eq:logtrans}. Since the equation is linear and translationally invariant, we can solve it by means of the standard Fourier transform.

We can express any perturbation as a linear combination of plane waves labelled by a wavenumber $\wn$,
\begin{equation}\label{eq:wave}
\epsilon_k(x,t)=e^{i(\wn x+\omega t)}.
\end{equation}
In terms of that wavenumber, we get the following non-local dispersion relation
\begin{widetext}
\begin{equation}\label{eq:lams}
\begin{split}
 i\omega(\wn )=&-u_0\int dy\ \xs(y)\left[e^{(\rho-i\wn) y}+e^{i\wn y}
 -\int dz e^{\rho(y+z)}\xg(y,z)\left(e^{i\wn z}+e^{-i\wn(y+z)}-1\right)\right]\\
 &+\int dy\int dz\ \xt(y,z)\Big[e^{-\rho y}(e^{i\wn y}-1)
 +e^{(\rho-1)z}(1-e^{-y})(e^{-i\wn z}-1)\Big]+i\wn\xa,
\end{split}
\end{equation}
where we have used the steady-state condition \eqref{eq:stecon1} to eliminate the mortality rate $\xb$.

The sign of $\lambda(k)=-{\rm Im}(\omega(k))$ determines stability. If $\lambda(k)$ is positive then the amplitude of the plane wave \eqref{eq:wave} with wavenumber $k$ grows
exponentially with time, rendering the steady state unstable. We find that $\lambda(k)=\lambda_{\rm\wp}(k)+\lambda_{\rm\wt}(k)$ where
\begin{equation}\label{eq:lamp}
 \lambda_{\rm\wp}(k)=u_0\int dy\ \xs(y)\left(-(e^{\rho y}+1)\cos(\wn y)
 +\int dz e^{\rho (y+z)}\xg(y,z)\left(\cos(\wn z)+\cos(\wn(y+z))-1\right)\right)
\end{equation}
and
\begin{equation}\label{eq:lamt}
 \lambda_{\rm\wt}(k)=\int dy\int dz\ \xt(y,z)\left(e^{-\rho y}(\cos(ky)-1)
 +e^{(\rho-1)z}(1-e^{-y})(\cos(kz)-1)\right).
\end{equation}
\end{widetext}
Note that the maintenance rate parameter $\xa$ and the death rate parameter $\xb$ no longer appear in these expressions.
Because a parent always uses weight during spawning, $y$ is positive wherever $b(y,z)$ is nonzero, and hence we can see that $\lambda_{\rm\wt}(\wn)$ is negative for any nonzero $k$.
This shows that reproduction always has a stabilising effect.

In the remainder of the section we will discuss the consequences on the stability of the steady state for the choices \eqref{eq:gaufeed}, \eqref{eq:genker} and \eqref{eq:t}
for the reproduction and predation functions. We show in Figure~\ref{fig:lamPB} the combined eigenvalue spectrum $\lambda(k)=\lambda_{\rm\wp}(k)+\lambda_{\rm\wt}(k)$ for two different values of $\rho$, corresponding to
both a stable ($\rho=0.9$, $\mu=-8$) and unstable ($\rho=1.1$, $\mu=-10$) fixed point $u_0$. We find that the
spectrum for $\rho = 0.9$ is everywhere negative, corresponding to a stable steady state. We can see in the inset that the spectrum for $\rho=1.1$ is positive for small wavenumbers, leading to an instability of the steady state against very long-wavelength perturbations.  At  higher $k$ the spectrum is more negative, indicating stronger stability against short-wavelength perturbations.

In Figure~\ref{fig:lamPbetaQ} we show the contribution from predation $\lambda_{\rm\wp}(k)$ for two different values of $\beta$ with $Q=0.9$ and for two different values of $Q$ for $\beta=5.5$.
Increasing $Q$ from the value of $0.2$ used in \cite{Datta2010} has a considerable stabilising effect. Note that we are allowed to increase $Q$ because in this work we have separated
out the losses due to maintenance processes. We can see as well that decreasing the preferred body size ratio between predator and prey has a stabilising effect. Nevertheless, for realistic values of the parameters, the contribution from predation alone is positive at some wavenumbers $k$, showing that reproduction is required to achieve stability.

In order to characterise the effect of variability in $Q$ we have considered non-zero $\sigma_{\psi}$ in Figure~\ref{fig:lamspsi}. We have to keep the standard deviation $\sigma_\psi$
sufficiently small so that the probability that the absorption efficiency is above $100\%$ is negligible. This leads us to impose that
\begin{equation}
\sigma_\psi \ll (1-Q)e^{-y}
\end{equation}
for some typical value $y$ of the log of the predator:prey mass ratio. In practice, typical values are $1-Q\approx 0.1$ and $y \approx 5$, so $\sigma_\psi\ll 6\cdot 10^{-4}$. Therefore
the fact that the predator:prey mass ratio is so large implies that $\sigma_\psi$ has to be very small, and to check its influence in stability we have chosen values around $10^{-4}$.
The effect is negligible for small wavenumbers, although become slightly appreciable for highly oscillating plane waves. In Figure~\ref{fig:lamspsi} we plot the difference
$\lambda_{\rm\wp}-\lambda_{\rm\wp}^{(0)}$, being $\lambda_{\rm\wp}^{(0)}$ the real part of the eigenvalue for $\sigma_{\psi}=0$. Although the effect is very small, the variability
in the feeding efficiency always enhances the stability of the steady state. In Appendix~\ref{s:absorption} we show that the effect of these small fluctuations in the feeding
efficiency consist on adding a diffusion term to the model.

We have also studied the contribution $\lambda_{\rm\wt}$ that reproduction makes to the eigenvalue spectrum for various values for $\mu$ and $\nu$. The results are shown in Figure~\ref{fig:lamBmunu}.
As explained earlier, the spectra are always negative, showing that reproduction has a stabilising effect. Asymptotically $\lambda_{\rm\wt}$ converges to a
negative constant. Variations in $\mu$ and $\nu$ affect the oscillatory behaviour found for small values of $k$.

\section{\label{s:discussion}Conclusions}

In this paper we made use of the fact that the power-law size spectrum that is observed in the pelagic ecosystem will be predicted by any dynamic model that is invariant under
scale transformations. That allowed us to generalise earlier models without spoiling the prediction of a power-law steady state. Where earlier models only included the effects
of predation and intrinsic mortality, we added terms modelling maintenance costs and reproduction and also allowed variability in the absorption efficiency. Inclusion of maintenance and reproduction has increased the stability of the steady-state solution.

We did not go into much ecological detail in this paper and made no attempt to determine the parameters of the model directly from ecological data.
Nevertheless, by exploiting scale invariance,
we made several observations that are of ecological relevance and that were not clearly made in previous work:

1. The power-law exponent for the size spectrum is fixed solely by the scaling properties of the parameters of the model. No detailed investigation of the model and its solutions is
required to determine it. The exponent does not depend on details like the preferred predator:prey mass ratio, the feeding efficiency, the variability in feeding behaviour, the
absorption efficiency, the maintenance costs, the mortality rate, or the details of reproduction.

This is in contrast to the results of earlier works in which only predation was considered. In that special case there are not enough scaling relations to fix the steady state exponent
and it will depend on the details of the model.

It is a crucial aspect of our model that it contains both processes that are density-dependent (predation) and processes that are density-independent (maintenance respiration,
intrinsic mortality, reproduction). It is the relative scaling of the rates for these processes that determines the steady-state power-law exponent.

Camacho and Sol\'e \cite{CamachoSole2001} studied the steady-state power-law exponent in a model with intrinsic mortality and reproduction. However they assumed that mortality and
reproduction rates were proportional to predation rates. Thus, in effect, all their processes were assumed to be density-dependent and again the steady-state exponent was
not determined by scaling arguments alone.

2. The assumption of scale invariance leads to predictions about the scaling behaviour \eqref{eq:rtx} of the various parameters of the model, and these can be tested through
observation, as discussed in Section~\ref{s:scale}.

We can also make a prediction that has not yet been tested. The prey selection function $\xs$ can be related to data on the stomach contents of fish as follows.
Let $l(x,x')dx'$ be the observed average number of prey with logweight between weight $x'$ and $x'+dx'$ found in the stomach of a fish of logweight $x$. This stomach content reflects
what prey a fish has been eating recently, which is determined by the same predation rates that we used in constructing our model. Thus we get (see equations \eqref{eq:psa}
and \eqref{eq:logtransp})
\begin{equation}
 l(x,x') = e^{(\rho-\xi)x}\xs(x-x') u(x') T,
\end{equation}
where $T$ is a time related to the speed of digestion.
We can assume that the fish density $u(x')$ is close to the steady-state density \eqref{eq:ssx}. Substituting the steady-state density $u(x')=u_0e^{-\rho x'}$ into the above equation
gives
\begin{equation}
 l(x,x') =  u_0e^{-\xi x}\xs(x-x') T.
\end{equation}
So scale invariance makes a prediction for the form of the stomach content data. There is a lot of data available \cite{Barnes2008} and it should be possible, via a careful analysis of
this data, to determine the scaling exponent $\xi$. We predict that this will confirm that $\xi\approx 1/4$.

3. From the condition that in the steady state the number of individuals must be conserved we deduced a relation between the parameters of the model which in particular restricts the exponent $\rho$ to an interval. We saw that for biologically reasonable values of the parameters this interval is around $\rho\approx 1$, which is in agreement with observations.

4. We have seen that the steady state in the model without reproduction, and with $\xi=0$, is unstable against perturbations in the overall population density $u_0$. In fact this was
our motivation for including a reproduction term in the model in this paper. We have derived the condition \eqref{eq:us} for the magnitude of the reproduction rate with respect to the
magnitude of the maintenance and mortality rates which ensures stability against this perturbation.

5. We have studied the stability of the steady state against all small perturbations in the case $\xi=0$. The stability is determined by the sign of $\lambda(k)$, which has the two
contributions $\lambda_{\rm\wp}(k)$ from predation and $\lambda_{\rm\wt}(k)$ from reproduction. Maintenance and mortality do not enter these expressions. The contribution
$\lambda_{\rm\wp}(k)$ coincides with that calculated in \cite{Datta2010} in the case of fixed absorption efficiency, however with the conversion efficiency $K$ replaced by the much
larger absorption efficiency $Q$. We have seen that this has an important stabilising effect. The contribution $\lambda_{\rm\wt}(k)$ from reproduction is always negative, thus
enhancing stability.

6. We have generalised previous models for the predation process to allow for variability in the absorption efficiency and have found that this does not have a big impact on the
stability of the steady state. In Appendix~\ref{s:absorption} we show that this small effect can also be modelled by an additional diffusion term in the equations, which leads to simpler expressions.

In future we intend to extract further interesting information from the model by taking more detailed ecological information into account. In particular we intend to use theoretical
results from metabolic theories about the energy budget \cite{Kooijman2000,Nisbet2000,Brown2004} to determine the relative strength of the various processes in our model and observational results to choose appropriate values for parameters.

The analytic derivation of solutions and study of their stability that we performed in this paper should be pushed further. Numerical studies of the jump-growth model \cite{Law2009}
has hinted at the existence of travelling wave solutions. It would be nice to learn more about them within the framework of scale-invariance, possibly by extending our analysis in
Appendix~\ref{s:sst}. We would also like to investigate the speed at which perturbations move through the size spectrum.

Our stability analysis was restricted to the case $\xi=0$. In this case the symmetry algebra $\Aff$, generated by time-translations and scale transformations, simplifies to the
abelian symmetry group of translations in $t$ and $x$. This allowed us to perform the stability analysis in terms of plane waves. It would be nice if the technique could be extended
to the non-abelian case, possibly using the techniques from non-commutative harmonic analysis \cite{Kirillov1994}.

The power-law size spectrum has been observed over many orders of magnitude and covers not only fish but also all types of plankton and even inanimate particles. Our model is
appropriate only for organisms that feed by swallowing smaller organisms and that reproduces by spawning a large number of smaller organisms. It is not appropriate for phytoplankton or inanimate particles.
Other models are needed for these and it will be interesting to see how these models can be
coupled together. We propose that again the guiding principle should be scale invariance.

There is a limit to the amount of detail that can be incorporated into
a community size spectrum model like the one described in this paper.
In particular, because only the size of individuals is taken into
account, their different behaviour in different life stages can not be
modelled. A first step in the direction of refining the model was
taken in \cite{AndersenBeyer2006} where an individual is described not
only by its size but also by one species-specific trait, namely its
size at maturity. That allowed individuals to follow more realistic
ontogenetic growth curves. An interesting observation of
\cite{AndersenBeyer2006} was how the size spectra for the different
species, each of which singles out a particular scale in the form of
the maturity size, combine into a power-law community spectrum
described by a scale-invariant McKendrick--von Foerster equation.

In \cite{AndersenBeyer2006} the size of offspring was encoded through
a boundary condition on the species' size spectrum but when combining
these spectra into a community size spectrum the boundary condition
had to be ignored in order to achieve a power-law steady-state size
spectrum. It would be interesting to see if that work could be
extended to include a realistic model of the reproduction process and
still produce a power-law community spectrum.

In a pure size spectrum model like ours, that does not take life
stages into account, a realistic modelling of reproduction is not
possible. For example, the offspring of many marine species start
their life in an egg stage during which they are not part of the
consumer spectrum but are already preyed upon. In our model we had to
neglect this fact. The scaling relation \eqref{eq:qsci} implies that
the typical size of offspring is proportional to the size of the
parent. However, because we can not model the egg life stage, we can
not claim that this gives a reliable prediction for the relation
between egg size and parent size. Indeed, while such a relation may
hold for copepods \cite{HuntleyLopez1992}, it certainly does not hold for
fish \cite{fishbase}, where most species lay eggs of a similar size.

One might be tempted to simply replace the reproduction function
\eqref{eq:qsci} by one that represents the fact that most fish lay
eggs of a similar size. However this would immediately lead to a
steady state solution that is not a power law. The steady state
spectrum would exhibit a peak at the preferred egg size which is not
observed in the size-spectrum data. The correct approach is not to
break the scale invariance of the community spectrum model, but
instead to study how more detailed models at the species level can
combine into a scale-invariant model at the community level, in the
spirit of \cite{AndersenBeyer2006,Hartvig2010}. The species-level
models would have to take into account effects like the duration of
the egg stage, the separation between spawning grounds and feeding
grounds, the various spawning strategies, like for example the laying
of eggs in clusters, and many more. How and why the combination of
many such detail-rich species-level models can lead to a
scale-invariant community-level model is an intriguing problem and it
is to be hoped that the experience that mathematical physics has with
the emergence of scale invariance in complex systems can be exploited.

\begin{acknowledgments}
We thank Richard Law and Jos\'e Cuesta for useful comments and discussions. J.A.C. acknowledges funding by
projects MOSAICO (Ministerio de Educaci\'on y Ciencia), MODELICO (Comunidad de Madrid), COST Action MP0801
(European Science Foundation) and by a contract from Comunidad de Madrid and Fondo Social Europeo.
\end{acknowledgments}

\appendix

\section{\label{s:master}Systematic expansion of the master equation}

Our model is a Markov model and can be described by a master equation that gives the time evolution of the probability $P({\bf n},t)$ that the system is in the state ${\bf n}$ at
time $t$. There will be a contribution from each of the processes involved,
\begin{equation}\label{eq:master}
 \frac{\partial P}{\partial t}=
 \left(\frac{\partial P}{\partial t}\right)_{\rm\wp}
 +\left(\frac{\partial P}{\partial t}\right)_{\rm\wt}
 +\left(\frac{\partial P}{\partial t}\right)_{\rm\wa}
 +\left(\frac{\partial P}{\partial t}\right)_{\rm\wb}.
\end{equation}
This appendix will be devoted to the systematic expansion of the master equation in powers of the inverse system volume $\V^{-1}$ \cite{vanKampen}.
This expansion separates the macroscopic behaviour from the fluctuations, and gives a
linear Fokker--Planck equation describing the fluctuations. Since the procedure is quite similar for the four processes we are considering, we will get into details
for the predation process and will give the contributions of the other processes thereafter. An alternative derivation of the deterministic, macroscopic equation can
be found in \cite{Datta2008}.

A concise way of writing the contribution of predation events to the master equation uses the step operator notation \cite{vanKampen}. Since the probability to undergo a predation event
in $dt$ is $\wp_{ijk} n_i n_j \V^{-1}$, we have
\begin{equation}\label{eq:mpredation}
\left(\frac{\partial P({\bf n},t)}{\partial t}\right)_{\rm\wp} = \sum_{i,j,k}\frac{\wp_{ijk}}{\V}(\mathbb{E}_i\mathbb{E}_j\mathbb{E}_k^{-1}-\mathbb{I})[n_i n_jP({\bf n},t)],
\end{equation}
where the step operators act on any function $f({\bf n})$ as $\mathbb{E}_i f({\bf n}) = f(\dots, n_i+1,\dots)$ and $\mathbb{E}_i^{-1} f({\bf n}) = f(\dots, n_i-1,\dots)$.

In Section~\ref{s:det} we already introduced the split of each variable $n_i(t)$ into a deterministic, macroscopic component $\phi_i(t)$ describing the density of individuals in
weight bracket $i$, and a fluctuation component $\eta_i(t)$ as
\begin{equation}\label{eq:kampena}
n_i(t) = \V\phi_i(t)+\V^{1/2}\eta_i(t).
\end{equation}
The new stochastic variables $\eta_i$ have a probability
distribution $\Pi({\boldsymbol \eta},t)=\V^{1/2}P({\bf n},t)$ and
\begin{equation}
\left(\frac{\partial P({\bf n},t)}{\partial t}\right)_{\rm\wp} = \V^{-1/2}\frac{\partial \Pi({\boldsymbol \eta},t)}{\partial t} -
\sum_i\frac{\partial \Pi({\boldsymbol \eta},t)}{\partial \eta_i}\frac{d\phi_i}{dt},
\end{equation}
where we have used that $\V^{-1/2}d\eta_i/dt = -d\phi_i/dt$ (this follows from \eqref{eq:kampen}, taking the time derivative for fixed $n_i$). The step operator $\mathbb{E}_i$ now
transforms $\eta_i$ to $\eta_i+\V^{-1/2}$, and can be expanded as
\begin{equation}
\mathbb{E}_i = \mathbb{I}+\V^{-1/2}\frac{\partial}{\partial\eta_i}
 +\frac{1}{2}\V^{-1}\frac{\partial^2}{\partial\eta_i^2}+\cdots
\end{equation}
Substituting this expansion into the master equation \eqref{eq:mpredation} we arrive at an equation containing different powers of the system volume $\V$. The highest order
($\V^0$) terms in
the expansion contain only macroscopic variables $\phi_i$ and vanish if these satisfy the macroscopic equation \eqref{eq:deterpred}.

Terms at next order ($\V^{-1/2}$) give a linear Fokker-Planck equation for the probability distribution $\Pi(\boldsymbol\eta)$ of the fluctuations,
\begin{equation}\label{eq:fokker}
\left(\frac{\partial\Pi}{\partial t}\right)_{\rm\wp}=
 -\sum_{ij}L^{\rm\wp}_{ij}\frac{\partial}{\partial\eta_i}\left(\eta_j\Pi\right)
 +\frac{1}{2}\sum_{ij}N^{\rm\wp}_{ij}\frac{\partial^2 \Pi}{\partial\eta_i\partial\eta_j}.
\end{equation}
By introducing the symmetric combination
\begin{equation}
 f_{ijk}=\frac{1}{2}(\wp_{ijk}+\wp_{jik}),
\end{equation}
we can give concise expressions for the coefficients in the Fokker--Planck equation,
\begin{equation}
\begin{split}
 L^{\rm\wp}_{ii}&=-2\sum_{jl}f_{ijl}\phi_j,\\
 L^{\rm\wp}_{ij}&=-2\sum_l\left(f_{ijl}\phi_i-f_{lji}\phi_l\right),\\
 N^{\rm\wp}_{ii}&=\sum_{jl}\left(2f_{ijl}\phi_i\phi_j+f_{jli}\phi_j\phi_l\right),\\
 N^{\rm\wp}_{ij}&=2\sum_l\left(f_{ijl}\phi_i\phi_j-f_{ilj}\phi_i\phi_l-f_{lji}\phi_l\phi_j\right).
\end{split}
\end{equation}
Terms at higher order in $\V^{-1}$ specify how the fluctuations deviate from being Gaussian. Fluctuations are seen to be damped by a factor $\V^{-1/2}$, and the volume $\V$ of our
system is very large. This justifies that we can concentrate on the study of the macroscopic equation in this paper.

The remaining processes involved in the model can be treated in a similar fashion. The master equation for reproduction can be written as
\begin{equation}\label{eq:mreprod}
\left(\frac{\partial P({\bf n},t)}{\partial t}\right)_{\rm\wt} =
\sum_{i,j,k}\wt_{ijk}(\mathbb{E}_i\mathbb{E}_j^{-1}\mathbb{E}_k^{-m_{ijk}}-\mathbb{I})[n_i P({\bf n},t)].
\end{equation}
At leading order the expansion in powers of $\V$ gives the deterministic contribution \eqref{eq:deterrep}. At next order the fluctuations are governed by a Fokker-Planck equation like \eqref{eq:fokker} but with coefficients given by
\begin{equation}
\begin{split}
 L^{\rm\wt}_{ii}&=-\sum_{jl}\wt_{ijl},\\
 L^{\rm\wt}_{ij}&=\sum_l\left(\wt_{ijl}+\wt_{jli}m_{jli}\right),\\
 N^{\rm\wt}_{ii}&=\sum_{jl}\left(\wt_{ijl}\phi_i+\wt_{jil}\phi_j+\wt_{jli}m_{jli}^2\phi_l\right),\\
 N^{\rm\wt}_{ij}&=\sum_l\left(-\wt_{ijl}\phi_i-\wt_{ilj}\phi_i+\wt_{lji}m_{lji}\phi_l\right).
\end{split}
\end{equation}

The effect of maintenance on the time evolution of the probability $P({\bf n},t)$ is given by
\begin{equation}\label{eq:mmain}
\left(\frac{\partial P({\bf n},t)}{\partial t}\right)_{\rm\wa} =
\sum_i \wa_i(\mathbb{E}_{i-1}^{-1}\mathbb{E}_i-\mathbb{I})[n_i P({\bf n},t)],
\end{equation}
which leads to \eqref{eq:detmain} at leading order and to the following contributions to the coefficients of the Fokker--Planck equation,
\begin{equation}
\begin{split}
 L^{\rm\wa}_{ii}&=-\wa_i,\\
 L^{\rm\wa}_{ij}&=\delta_{i+1,j}\wa_j,\\
 N^{\rm\wa}_{ii}&=\wa_i\phi_i+\wa_{i+1}\phi_{i+1},\\
 N^{\rm\wa}_{ij}&=-\delta_{i+1,j}\wa_j\phi_j-\delta_{i,j+1}\wa_i\phi_i.
\end{split}
\end{equation}

Finally, the contribution of intrinsic mortality to the master equation is
\begin{equation}\label{eq:mmor}
\left(\frac{\partial P({\bf n},t)}{\partial t}\right)_{\rm\wb} =
\sum_i \wb_i(\mathbb{E}_i-\mathbb{I})[n_i P({\bf n},t)],
\end{equation}
which gives \eqref{eq:detmor} and for the fluctuations we get contributions to the diagonal coefficients in the Fokker-Planck equation,
\begin{equation}
\begin{split}
 L^{\rm\wb}_{ii}&=-\wb_i,\\
 N^{\rm\wb}_{ii}&=\wb_i\phi_i,\\
\end{split}
\end{equation}
and the contributions to the off-diagonal coefficients are equal to zero. Note that the master equation for the complete model \eqref{eq:master} is simply the sum of all the
contributions, so the combined effect of all the processes in the macroscopic and the Fokker--Planck equation is just the sum of all the terms.

\section{\label{s:uzero}Positivity of $u_0$}

We are going to show that under very natural assumptions the steady-state population coefficient $u_0$ is positive when mortality is not too large and the steady-state exponent satisfies $\rho>1/e-\xi$. This inequality holds for the observed values that are in the region of $\rho\approx 1$ and $\xi\approx 1/4$. The assumptions are that
\begin{enumerate}
\item Predators only eat prey that are smaller than themselves. This means that $\xs(y)\neq 0$ only if $y>0$;
\item Predators always gain weight during feeding, i.e., $\xg(y,z)\neq 0$ only if $z>y$;
\item Offspring are always smaller than their parent and the parent always looses weight during spawning, i.e., $\xt(y,z)\neq 0$ only if both $y>0$ and $z<0$;
\item All the functions $\xs, \xg$ and $\xt$ are non-negative and have non-vanishing integrals.
\end{enumerate}

We define the functions $q_1(\rho)=c_{\rm\wp}+(\rho+\xi)f_{\rm\wp}$ and
$q_2(\rho)=c_{\rm\wt}-d+(\rho+\xi)f_{\rm\wt}$. These are the denominator and numerator in the expression \eqref{eq:u0} for $u_0$. We will show that these two functions are both positive under the above assumptions, which implies the positivity of $u_0$.

The derivative of $q_1$ with respect to $\rho$, keeping the other parameters constant, is
\begin{widetext}
\begin{equation}\label{cpr}
\frac{\partial q_1}{\partial \rho} = \int dy s(y)e^{\rho y}\left(y+(\rho+\xi-1)\int dz (y+z)e^{(\rho+\xi)z}\xg(y,z) \right)+f_{\rm\wp}.
\end{equation}
\end{widetext}
Assumptions 1 and 2 imply that, wherever the integrand is nonzero, we have $z>y>0$, hence $e^{(\rho+\xi)z}>1$ and
\begin{equation}
\int dz (y+z)e^{(\rho+\xi)z}\xg(y,z)>y,
\end{equation}
since $\xg$ is normalised to unity. Substituting this back into \eqref{cpr} and using assumption 4 shows that $\partial q_1/\partial \rho>0$ for all $\rho$.
Moreover, at the particular point $\rho=-\xi$ we can calculate
\begin{equation}
q_1(-\xi)=\int dy\ \xs(y)e^{-\xi y} > 0.
\end{equation}
Therefore the monotonicity of $q_1$ implies that $q_1(\rho)>0$ for all $\rho>-\xi$.

Similarly we calculate
\begin{widetext}
\begin{equation}
\frac{\partial q_2}{\partial \rho} = \int dy\int dz \, b(y,z)\left(1-ye^{-(\rho+\xi)y}+e^{-z}(1-e^{-y})(ze^{(\rho+\xi)z}+1)\right).
\end{equation}
\end{widetext}
According to assumption 3, $y>0$ and $z<0$ wherever the integrand is nonzero, therefore $1-e^{-y}>0$. For $\rho>1/e-\xi$ we also have $ye^{-(\rho+\xi)y}<1$ and $ze^{(\rho+\xi)z}>-1$ wherever the integrand is
nonzero. This, together with assumption 4, implies that $\partial q_2/\partial \rho>0$ and hence $q_2$ is strictly increasing with $\rho$. This implies the positivity of $q_2$ for all $\rho>1/e-\xi$ if $q_2$ is nonnegative at $\rho=1/e-\xi$, i.e., if
\begin{equation}\label{eq:mor}
\xb < c_{\rm\wt}|_{\rho=1/e-\xi}+\frac{1}{e}f_{\rm\wt}|_{\rho=1/e-\xi}.
\end{equation}
This represents a positive bound on $\xb$ because $f_{\rm\wt}$ is always positive and $c_{\rm\wt}>0$ at $\rho+\xi=1/e$.
In the particular case of absence of intrinsic mortality ($\xb=0$) the above restriction automatically holds. We have proven that $u_0$ is positive for $\rho>1/e-\xi$ if
\eqref{eq:mor} is true.

\section{\label{s:sst}General scale-invariant solution}

In this appendix we will look at solutions that are scale-invariant but not time-independent. They take the form
\begin{equation}
 u(x,t)=e^{-\rho x} f(\chi) \text{ where } \chi=x-\ln(t)/\xi
\end{equation}
for some function $f$. Substituting this Ansatz into the evolution equation \eqref{eq:logtrans} gives
\begin{widetext}
\begin{equation}
\begin{split}
 -\frac{1}{\xi}e^{\xi \chi }f'(\chi )&=\int dy\,s(y)\Big[-e^{\rho y}f(\chi )f(\chi -y)-e^{-\xi y}f(\chi )f(\chi +y)
 +\int dz\, \xg(y,z)e^{(\rho+\xi)z+\rho y}f(\chi -z)f(\chi -y-z)\Big]\\
 &+\int dy\int dz\,\xt(y,z)\Big[ -f(\chi )+e^{-(\rho+\xi) y}f(\chi +y)
 +e^{(1-\rho-\xi)z}(1-e^{-y})f(\chi +z)\Big]\\
 &+\xa \Big[-(\rho+\xi)f(\chi )+f'(\chi )\Big]-\xb\, f(\chi ).
\end{split}
\end{equation}
\end{widetext}
Unfortunately, this ordinary integro-differential equation for $f(\chi)$ is still difficult to solve in general.

An analytic solution can be found in the special case where only predation is considered. It is given by
\begin{equation}
 f(\chi)=f_0 e^{\xi \chi}.
\end{equation}
where the prefactor $f_0$ is determined by
\begin{equation}
1/f_0=\int dy\,s(y)\Big(e^{(\rho-\xi)y}+1-\int dz\,\xg(y,z)e^{(\rho-\xi)(y+z)}\Big).
\end{equation}
In terms of $u$ this solution reads
\begin{equation}
 u(x,t)=f_0e^{-(\rho-\xi)x}/t.
\end{equation}
Note how the exponents $\rho$ and $\xi$ only appear in the combination $\hat{\rho}=\rho-\xi$ whose value is determined by the scaling behaviour of the feeding function alone, see
\eqref{eq:rtx}.
The model without reproduction, maintenance and intrinsic mortality was treated in earlier work \cite{SilvertPlatt1978,BenoitRochet2004,Datta2008}, but this time-dependent
power-law solution is new.

\section{\label{s:absorption}Variability in absorption efficiency}

In the existing literature \cite{BenoitRochet2004,Datta2010}, the absorption efficiency was taken to be the same in all predation events. In Section~\ref{ss:param} we were allowing
variability in the absorption efficiency, modelled by a Gaussian distribution \eqref{eq:genker} with standard deviation $\sigma_\psi$.
The fact that the predator:prey mass ratio is so large implies that
$\sigma_\psi$ has to be very small ($\sigma_\psi\ll 6\cdot 10^{-4}$). This allows us to approximate the integral containing the Gaussian by Laplace's method. We will see that this will
lead to a diffusion term added to the model with fixed absorption efficiency.

With  $\xg$ chosen as in \eqref{eq:genker} the third term in the feeding part \eqref{eq:logtransp} of the model reads
\begin{multline}
F_3 = e^{(\rho-\xi)x}\int dy\ \xs(y)\int dz\ e^{-(\rho-\xi)z}g_{\sigma_\psi}(z-\psi(y))\\
u(x-z)u(x-y-z).
\end{multline}
After a shift in the integration variable $z$ this takes the form
\begin{equation}
F_3 = \frac{1}{\sqrt{2\pi}\sigma_\psi}\int dy\int dz h(x,y,z)e^{-z^2/2\sigma_\psi^2},
\end{equation}
where
\begin{multline}
h(x,y,z)=e^{(\rho-\xi)(x-z-\psi(y))}s(y)\\
u(x-z-\psi(y))u(x-y-z-\psi(y)).
\end{multline}
We can expand $h(x,y,z)$ in Taylor series of $z$ around $z=0$ up to second order and use Laplace's method (see e.g. \cite{BenderOrszag1999}) to evaluate the asymptotic behaviour of
the integral. Neglecting exponentially decaying terms, we can approximate
\begin{equation}
F_3\approx\int dy h(x,y,0)+\frac{\sigma_\psi^2}{4}\int dy\left[\frac{\partial^2}{\partial z^2}h(x,y,z)\right]_{z=0}+\cdots
\end{equation}
in the limit of $\sigma_\psi\ll 1$. Taking into account that
\begin{equation}
\frac{\partial^2}{\partial z^2} f(x-z) = \frac{\partial^2}{\partial x^2} f(x-z)
\end{equation}
holds for any sufficiently smooth function $f$, we finally get the asymptotic expansion of the model for $\sigma_\psi\ll 1$,
\begin{equation}\label{eq:smallsigmak}
\left(\frac{\partial u(x)}{\partial t}\right)_{\rm\wp}
 =-J(x)u(x)+\left(1+\frac{\sigma_\psi^2}{4}\frac{\partial^2}{\partial x^2}\right)E(x),
\end{equation}
where
\begin{equation}
J(x) = e^{(\rho-\xi)x}\int dy \xs(y)\left(u(x-y)+e^{(\rho-\xi)y}u(x+y)\right),
\end{equation}
\begin{multline}
E(x) = e^{(\rho-\xi)x}\int dy \xs(y)e^{-(\rho-\xi)\psi(y)}u(x-\psi(y))\\
u(x-y-\psi(y)).
\end{multline}
Therefore, allowing small fluctuations in the feeding efficiency has the effect of adding a diffusion term to the model.

\bibliography{simssd}

\end{document}